\documentclass[a4paper]{article}
\usepackage{RR}
\usepackage{hyperref}
\usepackage{amsfonts}
\usepackage{latexsym}
\usepackage{color}  %
\usepackage{pst-node} 
\usepackage{ae,aecompl,amsbsy,amssymb}
\usepackage{html} 
\newcommand{\coqdockw}[1]{\textsf{#1}}
\newcommand{\coqdocid}[1]{\textit{#1}}
\newcommand{\coqdoceol}{\setlength\parskip{0pt}\par}
\newcommand{\coqdocindent}[1]{\noindent\kern#1}
\newcommand{\coqdocmodule}[1]{\section*{Module #1}\markboth{Module #1}{}
}

\renewcommand{\backslash}{\mbox{\char`\\}~}

\RRdate{Novembre 2006}
\RRauthor{
Laurence Rideau
  \and
Laurent Théry
}
\authorhead{Rideau \& Théry}
\RRtitle{Formalisation des théorèmes de Sylow dans Coq}
\RRetitle{Formalising Sylow's theorems in Coq}
\titlehead{Sylow in Coq}

\RRresume{Ce rapport présente une formalisation des théorèmes de Sylow faite
dans le système {\sc Coq}. La formalisation s'est faite en deux semaines au
dessus de la librairie {\sc ssreflect} de Georges Gonthier.
Il y avait deux principales motivations pour formaliser les théorèmes de Sylow.
La première était de se familiariser avec la façon qu'a Georges de faire des preuves.
La seconde était de contribuer à l'effort collectif de formaliser un large ensemble
de la théorie des groupes en {\sc Coq}.
}
\RRabstract{This report presents a formalisation of Sylow's theorems done in {\sc Coq}.
The formalisation has been done in a couple of weeks on top of Georges
Gonthier's {\sc ssreflect}~\cite{ssreflect}. There were two ideas behind formalising Sylow's 
theorems. The first one was to get familiar with Georges way of doing proofs.
The second one was to contribute to the collective effort to formalise a large 
subset of group theory in {\sc Coq} with some non-trivial proofs.}

\RRmotcle{Th\'eorie des groupes, Théorème de Sylow, Formalisation des mathématiques} 
\RRkeyword{Group theory, Sylow's theorems, Formalisation of mathematics}
\RRprojet{Marelle}
\RRtheme{\THSym} 
\URSophia 
\begin{document}
\makeRT   

\section{Introduction}
Sylow's theorems are central in group theory. Any course has a section or a chapter
on them. Taking them as a first step in an effort to formalise group theory seemed
a good idea. One of these theorems is number 72 in the list of the 
100 theorems~\cite{T100} maintained by Freek Wiedijk. Surprisingly, only one
formalisation is known. It has been done in Isabelle by Florian Kammüller~\cite{Kam}.
The proof that has been formalised in Isabelle is due to Wielandt~\cite{Wielandt}.
It is a very concise and elegant proof. A central step in the proof is a
non-trivial combinatorial argument that is used to show the existence of a group with a particular property.
This is not the proof we have chosen to formalise. As we are interested in formalising Sylow's 
theorems not only as a mere exercise but as a base for further development, conciseness is nice but
reusability is much more important. We have chosen to follow the proof given by Gregory Constantine~\cite{Const}
in his group theory course.
It has the nice property of using one main tool, namely group actions,  
to prove most of the key results. The combinatorial argument that was present in the proof
of Wielandt is then reduced to a minimum. Most of our formalising time has then been spent proving theorems about 
groups not about numbers.

The presentation of this work is organised as follows. In a first section, we describe what we started
from. The main points we want to address are how {\sc ssreflect} is organised and how using this dedicated 
version of {\sc Coq} differs from using the standard one. 
In a second section, we outline the main steps of our proofs. Then, in a last section
we conclude.

\section{From types with decidable equality to finite types}

\subsection{Types with decidable equalitiy}
One of the key decision of {\sc ssreflect} is to base the development on objects not in {\tt Type} but
in {\tt eqType}, i.e objects for which equality is decidable.
{\small
\begin{verbatim}
Structure eqType : Type := EqType {
  sort :> Set;
    eq :  sort -> sort -> bool;
   eqP :  forall x y, reflect (x = y) (eq x y)
}.
\end{verbatim}}\noindent
{\tt eq} is the function that decides equality and {\tt eqP} the theorem that insures
that {\tt (eq x y)}, written in the following as {\tt  x == y}, is true iff {\tt x = y}. 
We call this the adequacy of equality.

Adding decidability on objects has the
nice consequence to equate the type {\tt bool}, the booleans, with the type {\tt Prop}, the propositions.
Of course, these two types are not identified since we are completely compatible with the standard way of
doing proofs in {\sc Coq}. Still, an inductive relation {\tt reflect} of type {\tt Prop -> bool -> Type}
holds all the information to coerce one into the other. 

In practice, booleans are always privileged with respect to propositions. For this, the coercion
{\tt is\_true} from booleans to propositions is used.
{\small
\begin{verbatim}
Coercion is_true b := b = true.
\end{verbatim}
}\noindent
As an example, let us consider equality and conjunction. Instead of stating a conjunction of two equalities
as {\tt x = y /\backslash z = t}, we prefer writing it using booleans as
{\tt x == y \&\& z == t}.
This simple modification gives a classical flavour  to the usually intuitionistic
prover {\sc Coq}. Moreover, proof scripts become more similar to the ones of other systems like {\sc Hol}.
In particular, as booleans accommodate the substitutivity property, rewriting becomes the tactic
number one. This reflection between {\tt bool} and {\tt Prop} is supported by the tactic
language with the so-called views. As an example, consider the reflection over conjunction
which is represented by the theorem {\tt andP }
{\small
\begin{verbatim}
Theorem andP: forall b1 b2 : bool, reflect (b1 /\ b2) (b1 && b2).
\end{verbatim}
}\noindent
Suppose now that we have to prove the following goal {\tt x == y \&\& z == t}.
In order to split this goal into two subgoals, we use a combination of two tactics: {\tt (apply/andP; split)}.
The first tactic converts the {\tt \&\&} into a {\tt /\backslash}, the second tactic can then perform
the splitting. Similarly for an hypothesis, if the goal is  {\tt x == y \&\& z == t -> A}
for an arbitrary {\tt A}, the tactic {\tt (move/andP; case)} performs the convertion and the destructuring.
Note that we can do even shorter combining view and case: {\tt case/andP}.

Some standard operations are defined on {\tt eqType}. For example, it is possible to build
the set of pairs of objects. The construction is the following:
{\small
\noindent\vskip5pt\noindent
{\tt Structure eq\_pair (d$_1$ d$_2$:\,eqType): Type := EqPair \{}
\noindent\vskip0pt\noindent
{\tt \ \   eq\_pi$_1$: d$_1$;} 
\noindent\vskip0pt\noindent
{\tt  \ \ eq\_pi$_2$: d$_2$}
\noindent\vskip0pt\noindent
{\tt \}.
\noindent\vskip5pt\noindent
{\tt Definition pair\_eq (d$_1$ d$_2$:\,eqType)  (u v:\,eq\_pair d$_1$ d$_2$):\,bool:=}
\noindent\vskip0pt\noindent
{\tt \quad let EqPair x$_1$ x$_2$:= u in}
\noindent\vskip0pt\noindent
{\tt \quad let EqPair y$_1$ y$_2$:= v in
 (x$_1$ == y$_1$) \&\& (x$_2$ == y$_2$).}}\noindent\vskip5pt\noindent}Once the adequacy of the equality is proved, we can build the expected type with decidable equality. 
This is represented by
the function {\tt prod\_eqType} with the following type
{\small\tt  prod\_eqType: eqType -> eqType -> eqType}.
\noindent

\subsection{Sets}
Sets are represented by their indicator function:
{\small
\begin{verbatim}
Definition set (d: eqType) := d -> bool.
\end{verbatim}
}\noindent
For example, the constructor of a singleton is defined as
\noindent\vskip5pt\noindent
{\small
{\tt Definition set1 x := fun y => (y == x).
}
}
\noindent\vskip5pt\noindent
A key construction is the one that allows to build a type {\tt d$_1$} with decidable equality from a set {\tt A}
whose carrier is a type {\tt d} with decidable equality. This is done using the constructor {\tt sub\_eqType}:
{\small
\noindent\vskip5pt\noindent
 {\tt sub\_eqType: forall d: eqType, set d -> eqType.}
\noindent\vskip5pt\noindent}{\tt d$_1$} is then {\tt (sub\_eqType d {\tt A})} and 
elements of {\tt d$_1$} are composed of elements of {\tt d} and a proof that they belong to {\tt A}.
{\small
\begin{verbatim}
Structure eq_sig (d: eqType) (A: set d): Set :=  EqSig {
   val: d; 
  valP: A val
}.
\end{verbatim}}
\noindent
Equality then only checks the first elements of the two records.
As sets are represented as indicators, this equality is adequate (there is only one
proof of ${\tt x} = {\tt true}$).  Over sets, there is also the usual extensional equality, 
i.e. {\tt A$_1$  =$_1$ A$_2$} iff {\tt A$_1$ x  ==  A$_2$ x} for all {\tt x}.
 
\subsection{Sequence}

Sequences are represented in a standard way
{\small 
\begin{verbatim}
Inductive seq (d: eqType): Type := Seq0 | Adds (x : d) (s : seq d).
\end{verbatim}}\noindent
Sequences are equipped with all the basic operations. In the following, we are
going to use two of these operations: {\tt size}, {\tt count}.
{\tt size} gives the number of elements of a sequence.
{\tt count} returns the number of elements of a set inside a sequence.

\subsection{Finite type}

The last construction before defining groups is the one
for creating finite types. A finite type is composed of
a type {\tt sort} with decidable equality, its sequence of elements and a proof that the sequence
contains each element of {\tt sort} once and only once.
{\small
\begin{verbatim}
Structure finType: Type := FinSet {
   sort :> eqType;
   enum :  seq sort;
  enumP :  forall x, count (set1 x) enum = 1
}.
\end{verbatim}}
\noindent
Note that this encoding of finite sets gives for free an order on the elements
of the finite set, i.e. the index of its occurrence in the sequence. The cardinality 
of a set {\tt A} over a finite type {\tt S} is defined as {\tt (count A (enum  S))}.
 It is written in the following as {\tt (card A)}.
 
\section{From finite groups to Sylow's theorems}

\subsection{Finite group, coset and subgroup}

A finite group contains a finite set, an unit element, an inverse function and
a multiplication with the usual properties.
{\small
\begin{verbatim}
Structure finGroup : Type := Finite {
  element:> finType;
     unit:  element;
      inv:  element -> element;
      mul:  element -> element -> element;
    unitP:  forall x, mul unit x = x;
     invP:  forall x, mul (inv x) x = unit;
     mulP:  forall x1 x2 x3, mul x1 (mul x2 x3) = mul (mul x1 x2) x3
}.
\end{verbatim}
}
\noindent
Given a multiplicative finite group {\tt G} and {\tt x}, {\tt y} two elements of {\tt G}, {\tt 1} is encoded
as {\tt (unit {\tt G})}, {\tt x$^{-1}$} as {\tt (inv G x)}, and  {\tt xy} as {\tt (mul G  x y)}.
Given a finite group {\tt G}, a set {\tt H} of {\tt G} and an element {\tt a} of {\tt G}, the left coset {\tt aH} (the right
coset {\tt Ha}) is the set of the elements {\tt ax} (respectively the set of elements {\tt xa})
 for all {\tt x} in {\tt H}. As we have {\tt x} in {\tt aH} iff {\tt a$^{-1}$x} is in {\tt H} (respectively
 {\tt x} in {\tt Ha} iff {\tt xa$^{-1}$} is in {\tt H}), we have the following definitions:
{\small
\noindent\vskip5pt\noindent
{\tt Definition lcoset H a: set G := fun x => H (a$^{-1}$x).}
\noindent\vskip0pt\noindent
{\tt Definition rcoset H a: set G := fun x => H (xa$^{^{-1}}$).}}
\noindent\vskip5pt\noindent
The function {\tt x $\mapsto$ ax} is a bijection between {\tt H} and {\tt aH}, so both
sets have same cardinality. Furthermore, every coset {\tt aH}
can be represented by a canonical element $\overline{{\tt a}}$ such that ${\tt aH}\,\, \hbox{\tt =$_1$}\,\, {\tt bH}$
iff $\overline{{\tt a}}\,\, \hbox{\tt ==}\,\, \overline{{\tt b}}$. Technically,
$\overline{{\tt a}}$ is encoded as {\tt (root (lcoset {\tt H}) {\tt a})}, which is the first element
in the sequence of the finite set that belongs to {\tt aH}. 

Subgroups are not defined as structures but as sets.
Their definition  is a bit intricate. The idea is to say that a set
{\tt H} is a subgroup if it is not empty, and if {\tt x} and {\tt y} are in {\tt H} so is
{\tt xy$^{-1}$}. This is sufficient. Since if {\tt H} is non empty, it contains at least an element
{\tt z}, so we have {\tt zz$^{-1}$ == 1} belongs to {\tt H}. Also, for all {\tt x} in {\tt H}, {\tt 1x$^{-1}$ 
 == x$^{-1}$}
also belongs to {\tt H}. Finally, if {\tt x} and {\tt y} belongs to {\tt H}, we have {\tt y$^{-1}$}
belongs to {\tt H}, so is {\tt x(y$^{-1}$)$^{-1}$  == xy}. In our definition,
{\tt 1} is used as a witness of non-emptiness. For the second condition, 
we rewrite it as ``if {\tt x} is in {\tt H} then {\tt H} is included in {\tt Hx}".
{\small 
\begin{verbatim}
Definition subgrp H := 
  H 1 && subset H (fun x => subset H (rcoset H x)).
\end{verbatim}}
\noindent
where {\tt (subset H$_1$ H$_2$)} is true iff for all {\tt x} in {\tt H$_1$}, {\tt x} is also in {\tt H$_2$}. 
In this definition, {\tt G} is given implicitly since the type of {\tt H} is {\tt (set {\tt G})}.
This definition is of little use for proving that a set is a subgroup.
As we are in a finite setting, a much more practical characterisation of a subgroup is that
it is a non-empty set that is stable by multiplication. This is represented in
our development by the theorem {\tt finstbl\_sbgrp}:
{\small 
\begin{verbatim}
Lemma finstbl_sbgrp: forall G (H : set G) (a : G),
       H a -> (forall x y, H x -> H y -> H (xy)) -> subgrp H.
\end{verbatim}}
\noindent

If {\tt H} is a subgroup, its left cosets partition {\tt G}: if {\tt z} is in the intersection {\tt aH} and {\tt bH},
there exist {\tt h$_1$} and {\tt h$_2$} such that {\tt ah$_1$  == z  == b h$_2$}, 
we get {\tt a  == b(h$_2$h$_1$$^{-1}$)} and {\tt b == a(h$_1$h$_2$$^{-1})$}, 
so {\tt aH  =$_1$  bH}. 
We denote {\tt (lindex {\tt H})} the number of canonical elements. We then get
that {\tt card {\tt G} = lindex {\tt H} * card {\tt H}}. As in our development
groups and subgroups differ in nature, groups hold the carrier while subgroups are only indicators, it is 
preferable to state Lagrange's theorem at the level of subgroups:
{\small 
\begin{verbatim}
Theorem lLaGrange: 
 forall G (H K: set G),
 subgrp H -> subgrp K -> subset H K => card H * lindex H K = card K.
\end{verbatim}}
\noindent
Now, {\tt lindex H K} denotes the number of coset of {\tt H} with respect to {\tt K}.
Note that we can always get back to the usual statement, using the fact that
 {\tt G} is a subgroup of itself.
\subsection{Conjugate, normaliser and normal subgroub}
Normal subgroups are needed for the proof of Sylow's theorem. In order to 
define them, we first define the conjugate operation.
{\small
\noindent\vskip5pt\noindent
{\tt Definition y$^{\tt x}$ := x$^{-1}$yx}.
\noindent\vskip5pt\noindent
}\noindent
Then, given an arbitrary element {\tt x} and an arbitrary set {\tt H}
the conjugate set {\tt xHx$^{-1}$} is defined as follows:
{\small
\noindent\vskip5pt\noindent
{\tt Definition conjsg H x := fun y => H y$^{\tt x}$.
\noindent\vskip5pt\noindent
}\noindent}
{\tt y} is in {\tt xHx$^{-1}$} iff {\tt x$^{-1}$yx} is in {\tt H}.
We are now ready to define the notion of normal subgroup. {\tt H}
is normal in {\tt K} iff for all element {\tt x} in {\tt K},
{\tt xHx$^{-1}$ =$_1$ H}. It is in fact sufficient to require
that {\tt H} is included in {\tt xHx$^{-1}$} as both sets have same
cardinality. This gives the following definition:
{\small
\begin{verbatim}
Definition normal H K := subset K (fun x => subset H (conjsg H x)).
\end{verbatim}
}\noindent
Later in the proof of the first Sylow's theorem we use the property
that the quotient of a group by a normal subgroup is a group. This
is a direct consequence of normality that imposes that the operation
of the group behaves well with respect to cosets. The quotient group
is represented in our development by the group {\tt RG} composed with 
the roots of {\tt G} with respect to the left coset relation.

Given a subgroup {\tt H}, it is possible to build its normaliser, the set
of all {\tt x} in {\tt K} such that {\tt xHx$^{-1}$ = H} as:
{\small
\noindent\vskip5pt\noindent
{\tt Definition normaliser H K x :=}
\noindent\vskip0pt\noindent
{\tt \quad
  (subset K (fun z => (conjsg H x z == H z))) \&\& K x.}
\noindent\vskip5pt\noindent
}\noindent
By definition, we have that {\tt H} is normal in {\tt (normaliser H K)}.
This is the theorem {\tt normaliser\_normal}:
{\small
\begin{verbatim}
Lemma normaliser_normal: 
  forall G (H K : set G), subset H K -> normal H (normaliser H K).
\end{verbatim}}
       
\subsection{Group actions}

Group actions are the key construction for our final theorems. To define
an action, we need a group {\tt G}, a subgroup {\tt H} and a finite set
{\tt S}. This is written in our development as: 
{\small
\begin{verbatim}
Variable G : finGroup.
Variable H : set G.
Hypothesis sgrp_H: subgrp H.
Variable S : finType.
\end{verbatim}}\noindent
An action {\tt to} is a homomorphism
from {\tt H} to the permutations of {\tt S} (the bijections from {\tt S} to {\tt S}).
This is defined as:
{\small
\begin{verbatim}
Variable to: G -> (S -> S).
Hypothesis to_bij: forall x, H x -> bijective (to x).
Hypothesis to_morph: forall (x y: G) z, 
  H x -> H y -> to (xy) z = to x (to y z).
\end{verbatim}}\noindent
where the predicate {\tt bijective} indicates that the function is a bijection. Note that
we have arbitrary chosen to define our action {\tt to} on {\tt G} and only require
the properties of homomorphism and permutation to hold for elements of {\tt H}.

For an element {\tt a} of {\tt S}, we define its orbit as all the elements of {\tt S}
that can be reached from {\tt a} by the function {\tt to}. In other words, it is the
image of {\tt H} by the function that given an {\tt x} in {\tt G} associates {\tt (to x a)}.
{\small
\begin{verbatim}
Definition orbit a := image (fun x => to x a) H.
\end{verbatim}}\noindent
We can partition {\tt S} using the orbits. A key property of group action comes with the notion
of stabiliser. Given an element {\tt a} of {\tt S}, we call its stabiliser the set of
all the elements {\tt x} of {\tt H} that leave {\tt a} unchanged by the function {\tt to x}. Formally, this gives
{\small
\noindent\vskip5pt\noindent
{\tt Definition stabiliser a := fun x => (((to x a) == a)) \&\& (H x)).}
\noindent\vskip5pt\noindent
}\noindent
The stabiliser is clearly a subgroup of {\tt H} but the key property is that the
cardinal of the orbit of {\tt a}
and the index of the stabiliser of {\tt a} are equal.
{\small
\begin{verbatim}
Lemma card_orbit: forall a, card (orbit a) = lindex (stabiliser a) H.
\end{verbatim}}\noindent
to see this we just have to notice that we have {\tt (to x a) =$_d$ (to y a)} iff {\tt x$^{-1}$y} is in
{\tt (stabiliser a)}. For this, we write {\tt (to y a)} as {\tt (to x (to  (x$^{-1}$y) a))}) and use
the fact that {\tt to} is injective.

In the particular case where {\tt H} has cardinality $p^\alpha$ with $p$ prime, 
as orbits partition {\tt S} and their cardinality is an index, Lagrange's theorem gives us that 
these orbits are of cardinality $p^{\beta}$ with $\beta \le \alpha$. Now, if
we collect in the set {\tt S$_0$}
all the elements of {\tt S} whose orbit has cardinality $1=p^0$, i.e elements that are in the stabiliser of
every element of {\tt H}: 
{\small \tt
\noindent\vskip5pt\noindent
Definition S$_0$ a := subset H (stabiliser a).
\noindent\vskip5pt\noindent
}\noindent
we get our central lemma
{\small \tt
\noindent\vskip5pt\noindent
Lemma mpl: (card S) \% p = (card S$_0$) \% p.
\noindent\vskip5pt\noindent
}\noindent
where {\tt \%} is the usual modulo operation. All the orbits of cardinality $p^\beta$ with $0 <\beta \le \alpha$
cancel out in the modulo.
\subsection{Cauchy's theorem}
The proof of the first Sylow theorem is an inductive proof. Cauchy's theorem solves the base case.
This theorem states that if a prime $p$ divides the cardinality of a group, then there exists a subgroup
of cardinality $p$. More precisely, there exists an element {\tt a}, such that its cyclic group, i.e.
the set of all the {\tt  a$^i$}, is of cardinality $p$. As we did for Lagrange's, we state this theorem
at the level of subgroups. We take {\tt H} a subgroup of {\tt G} and a prime $p$ that divides the
cardinality of {\tt H}. We first consider {\tt H$^{p-1}$} the cartesian product $\underbrace{\hbox{\tt H} \times \dots \times \hbox{\tt H} }_{p-1}$. An element {\tt x} of {\tt H$^{p-1}$} is written as {\tt (h$_0$, \dots, h$_{p-2}$)}.
We have {\tt (card H$^{p -1}$) = (card H)$^{p-1}$}. We define $H^*$ a subset of {\tt H$^{p}$} as the image 
of {\tt H$^{p-1}$} by the function\\
\hbox{\qquad {\tt (h$_0$, \dots, h$_{p-2}$)} $\mapsto$
{\tt ($(\prod_{i=0}^{p-2} \hbox{h$_i$})^{-1}$, h$_0$, \dots, h$_{p-2}$)}}.\\
Clearly, we have  {\tt (card H$^*$) = (card H)$^{p-1}$} and every element {\tt (h$_0$, \dots, h$_{p-1}$)}
of {\tt H$^{p}$} such that $\prod_{i=0}^{p-1} \hbox{h$_i$} = 1$ is in {\tt H$^*$}.
Now we consider the additive group ${\mathbb{Z}_p}$ and the action {\tt to} from  ${\mathbb{Z}_p}$ to {\tt H$^*$} defined
as \\
\hbox{\qquad \tt n  $\mapsto$ \{ (h$_0$, h$_1$\dots, h$_{p-1}$)  $\mapsto$  (h$_{(0 +n)\% p}$, h$_{(1 +n)\% p}$,\dots, h$_{(p - 1 +n)\% p}$)\}}\\
Now, if we look at the set {\tt S$_0$} of the elements of orbit with cardinality 1. We can easily prove that {\tt S$_0$} is composed of the elements 
{\tt (h, \dots, h)} such that {\tt h$^p$ = 1}. In one direction, such elements clearly belong to {\tt S$_0$} since they are left unchanged
by any permutation of indexes.
Conversely, if an element {\tt x} belongs to {\tt S$_0$}, in particular {\tt (to 1 x)} is equal to {\tt x}. 
So, if we write {\tt x} as {\tt (h$_0$, \dots, h$_{p-1}$)}, this means {\tt (h$_0$, \dots, h$_{p-1}$)} is equal to
{\tt (h$_1$, \dots, h$_{0}$)} which in turn implies that {\tt h$_0$} is equal to {\tt h$_1$}, {\tt h$_1$} is equal to {\tt h$_2$}
and so on. Now, the {\tt mpl} lemma tells us that {\tt (card H$^*$) \% p = (card S$_0$) \% p}, but the cardinality
of {\tt H$^*$} is divisible by $p$ so we can conclude that the cardinality of {\tt S$_0$} is also divisible by $p$. 
As, $p \ge 2$, this means that there exists at least one element {\tt a} different from {\tt 1} in {\tt S$_0$}. For this element, 
we have $\hbox{\tt a}^{p} = 1$.
We have that the cardinality of the cyclic group of {\tt a} divides $p$ but as $p$ is prime and
{\tt a} is different of 1, the cardinality of its cyclic group is then exactly $p$. The exact statement of Cauchy's 
theorem in our development is 
{\small \tt
\noindent\vskip5pt\noindent
Theorem cauchy: forall G, (H : set G) p, 
\noindent\vskip0pt\noindent
\qquad  subgrp H -> prime p ->  p | (card h) -> 
\noindent\vskip0pt\noindent
\qquad    exists a, H a \&\& (card (cyclic a) == p).
\noindent\vskip5pt\noindent
}\noindent
where {\tt |} denotes the divisibility and {\tt cyclic} builds the cyclic group of an element.

\subsection{Sylow's theorems}

The first Sylow theorem tells us that if {\tt G} is a group and {\tt K} is a subgroup of {\tt G} of
cardinality $p^n s$ with $p$ prime and $p$, $s$ relatively prime, then there exists a subgroup of
{\tt K} of cardinality $p^n$. Such a subgroup of maximal cardinality in $p$ is called a Sylow $p$ subgroup.
It is defined in our development as
{\small \tt
\noindent\vskip5pt\noindent
Definition sylow K p H:= 
\noindent\vskip0pt\noindent
\quad  subgrpb H \&\& subset H K \&\& card H == expn p (dlogn p (card K)).
\noindent\vskip5pt\noindent
}\noindent
where {\tt expn} is the exponential function and {\tt dlogn} is the divisor logarithm, i.e
{\tt (dlogn p u)} is the maximal power of {\tt p} that divides {\tt u}.

The proof of the first Sylow theorem is done by induction. We are going to prove that
for all $i$, $0< i \le n$, there exists a subgroup of cardinality $p^i$. For $i=1$, the
existence is given by Cauchy's theorem. Now, suppose that there exists a subgroup {\tt H} of
cardinality $p^i$, we are going to prove that there exists a subgroup {\tt L} of cardinality
$p^{i+1}$. We are acting by left translation with {\tt H} on the left cosets of {\tt H} with respect
to {\tt K} as follows: \\
\hbox{\qquad \tt x  $\mapsto$ \{ yH  $\mapsto$  (xy)H \}}\\
The {\tt mpl} lemma gives us {\tt (card S$_0$) \% p = (lindex H K) \% p}.
But by Lagrange's theorem we know that {\tt (lindex H K)} is equal to $p^{n -i}s$.
As $i< n$, we can conclude that the cardinal of {\tt S$_0$} is divisible by $p$.
Now, if we look at the cosets that are in {\tt S$_0$}. They are the {\tt yH} such that
{\tt (xy)H = yH} for all {\tt x} in {\tt H}. This corresponds to {\tt y$^{-1}$Hy = H} so
{\tt y} is in {\tt (normaliser H K)}. So, we can deduce that {\tt (card S$_0$) =  (lindex H (normaliser H K))}.
This means that if we take the quotient of the normaliser {\tt (normaliser H K)} by {\tt H}, this is 
a group ({\tt H} is normal in its normaliser) and its cardinality which is {\tt (lindex H (normaliser H K))}
is divisible by $p$. We can then apply Cauchy's theorem and get the existence of a subgroup {\tt L$_1$} of
cardinality $p$ in the quotient. Taking the inverse image of {\tt L$_1$} by the quotient operation, we get a subgroup
{\tt L} of {\tt G} whose cardinality is {\tt card L$_1$ * card H = p p$^{i}$ = p$^{i + 1}$}. This ends
the proof of the first Sylow theorem. The exact formal statement of this theorem is the following:
{\small
\begin{verbatim}
Theorem sylow1_cor: forall G (K: set G) p,
       subgrp K -> prime p -> 0 < dlogn p (card K) -> 
       exists H : set G, sylow K p H.
\end{verbatim}}\noindent

The second Sylow theorem says that two Sylow $p$ subgroups {\tt L$_1$} and {\tt L$_2$} of {\tt K} are conjugate.
For the proof, we act by left translation with {\tt L$_2$} on the left coset of {\tt L$_1$}.
By the {\tt mpl} lemma, we know the {\tt (card S$_0$) \% p = (lindex L$_1$ K) \% p}. As {\tt L$_1$} is a Sylow $p$
group, we have by Lagrange's theorem that {\tt (lindex L$_1$ K)} is equal to  $s$, so is not divisible by $p$.
This means that {\tt (card S$_0$)} is not divisible by $p$, so there exists an {\tt x} in {\tt K} such that {\tt xL$_1$} is
in  {\tt S$_0$}. But for this {\tt x}, we know that for all {\tt y} in {\tt L$_2$}, {\tt (yx)L$_1$ = xL$_1$}, this means
that {\tt L$_2$} is included in {\tt xL$_1{\tt x}^{-1}$}. As both sets have same cardinality, we have  {\tt L$_2$ =$_1$ xL$_1{\tt x}^{-1}$}.
The exact formal statement of this theorem is the following: 
{\small \tt
\noindent\vskip5pt\noindent
Theorem sylow2\_cor: forall G (K: set G) p L$_1$ L$_2$,
\noindent\vskip0pt\noindent
\qquad       subgrp K -> prime p -> 0 < dlogn p (card K) -> 
\noindent\vskip0pt\noindent
\qquad       sylow K p L$_1$ -> sylow K p L$_2$ -> 
\noindent\vskip0pt\noindent
\quad\qquad         exists x : G, K x /\backslash L$_2$ =$_{1}$ conjsg L$_1$ x.
\noindent\vskip5pt\noindent
}\noindent

The third Sylow theorem  gives an indication on the number of Sylow $p$ groups. It says that this number
divides the cardinality of {\tt K} and is equal to 1 modulo $p$. In order to count the number of Sylow $p$
subgroup, we have to define the sylow subset of the power set of {\tt G} as:
{\small
\begin{verbatim}
Definition syset K p := fun (H: powerSet G) => sylow K p (subdE H).
\end{verbatim}}\noindent
Now, the first part of the third theorem that regards divisibility is proved acting with {\tt K}
on {\tt (syset K p)} as follows:\\
\hbox{\qquad \tt x  $\mapsto$ \{ L  $\mapsto$  xLx$^{-1}$ \}}\\
The second theorem tells us that all the elements of {\tt (syset K p)} are conjugate. So, from
one Sylow $p$ subgroup {\tt L} we can reach any other by conjugation. This
means that {\tt (syset K p)} contains one single orbit. So, {\tt (card (syset K p)) = (card (orbit L))}.
The theorem {\tt card\_orbit}
tells us the {\tt card (orbit L)} is equal to {\tt (lindex (stabiliser L) K)}.
Using Lagrange's theorem, we get that it divides {\tt (card K)}. The formal statement
of the first part of the third Sylow theorem is the following:
{\small
\begin{verbatim}
Theorem sylow3_div: forall G (K: set G) p,
       subgrp K -> prime p -> 0 < dlogn p (card k) -> 
       (card (syset K p)) | (card K).
\end{verbatim}}
For the second part, we consider {\tt H} a Sylow $p$ group for {\tt K}. We act with {\tt H}
on {\tt (syset K p)} by conjugation as before:\\
\hbox{\qquad \tt x  $\mapsto$ \{ L  $\mapsto$  xLx$^{-1}$ \}}\\
An element L is in {\tt S$_0$} if {\tt xLx$^{-1}$ =$_1$ L} for all {\tt x} in {\tt H}.
This means that {\tt H} is included in {\tt (normaliser L K)}. As we have {\tt (sylow K p H)},
we have also {\tt (sylow (normaliser L K) p H)}. This  holds also for {\tt L}, so we have 
{\tt (sylow (normaliser L K) p L)}. The second theorem tells us that {\tt H} and {\tt L} are
then conjugate in {\tt  (normaliser L K)}. But as {\tt L} is normal in its normaliser, this
implies that {\tt H =$_{1}$ L}. So {\tt (card S$_0$)} is equal to 1. If we apply the {\tt mpl}
lemma we get the expected result. The formal statement
of the second part of the third Sylow theorem is the following:
{\small
\begin{verbatim}
Theorem sylow3_mod: forall G (K: set G) p,
       subgrp K -> prime p -> 0 < dlogn p (card k) -> 
       (card (syset K p)) % p = 1.
\end{verbatim}}\noindent

\section{Conclusion}

Formalising Sylow's theorems has been surprisingly smooth. One reason has to do with
the fact that we have built our development on top of {\sc ssreflect}. This base
was used by Georges Gonthier for his proof of the four colour theorem. 
It  has already been tested on a large development, so it is quite complete. The only basic 
construction we had to add is the power set. Another reason that made our life simpler
is that we were working in a decidable fragment of the {\sc Coq} logic. No philosophical
issue about constructiveness slowed down our formalisation. Finally, Gregory Constantine's proof 
was perfect for our formalisation work. The only part of the formalisation that was ad-hoc
was the construction of the set {\tt H$^*$}. It represents only 360 lines of the 3550 lines
of the formalisation. The fact that this experiment was positive is clearly a good sign
for further formalisations in group theory.

\bibliographystyle{plain}
\bibliography{sylow}

\coqdocmodule{groups}

\medskip
\noindent
\coqdocid{Structure} \coqdocid{finGroup}: \coqdocid{Type}:= \coqdocid{Finite} \{\coqdoceol
\coqdocindent{1.00em}
\coqdocid{element}:> \coqdocid{finType};\coqdoceol
\coqdocindent{1.00em}\coqdocid{unit}\coqdocindent{1.70em}
: \coqdocid{element};\coqdoceol
\coqdocindent{1.00em}
\coqdocid{inv}\coqdocindent{2.10em}: \coqdocid{element} \ensuremath{\rightarrow} \coqdocid{element};\coqdoceol
\coqdocindent{1.00em}
\coqdocid{mul}\coqdocindent{1.90em}: \coqdocid{element} \ensuremath{\rightarrow} \coqdocid{element} \ensuremath{\rightarrow} \coqdocid{element};\coqdoceol
\coqdocindent{1.00em}
\coqdocid{unitP}\coqdocindent{1.05em}: \ensuremath{\forall}\coqdocid{x}, \coqdocid{mul} \coqdocid{unit} \coqdocid{x} = \coqdocid{x};\coqdoceol
\coqdocindent{1.00em}
\coqdocid{invP}\coqdocindent{1.50em}: \ensuremath{\forall}\coqdocid{x}, \coqdocid{mul} (\coqdocid{inv} \coqdocid{x}) \coqdocid{x} = \coqdocid{unit};\coqdoceol
\coqdocindent{1.00em}
\coqdocid{mulP}\coqdocindent{1.20em}: \ensuremath{\forall}\coqdocid{x$_1$} \coqdocid{x$_2$} \coqdocid{x$_3$}, \coqdocid{mul} \coqdocid{x$_1$} (\coqdocid{mul} \coqdocid{x$_2$} \coqdocid{x$_3$}) = \coqdocid{mul} (\coqdocid{mul} \coqdocid{x$_1$} \coqdocid{x$_2$}) \coqdocid{x$_3$}\coqdoceol
\noindent
\}.\coqdoceol

\medskip
\noindent
\coqdockw{Section} \coqdocid{GroupIdentities}.\coqdoceol

\medskip
\noindent
\coqdockw{Variable} \coqdocid{G}: \coqdocid{finGroup}.\coqdoceol

\medskip
\noindent
\coqdockw{Lemma} \coqdocid{mulgA}: \ensuremath{\forall}\coqdocid{x$_1$} \coqdocid{x$_2$} \coqdocid{x$_3$}: \coqdocid{G}, \coqdocid{x$_1$} \ensuremath{\times} (\coqdocid{x$_2$} \ensuremath{\times} \coqdocid{x$_3$}) = \coqdocid{x$_1$} \ensuremath{\times} \coqdocid{x$_2$} \ensuremath{\times} \coqdocid{x$_3$}.\coqdoceol

\medskip
\noindent
\coqdockw{Lemma} \coqdocid{mul1g}: \ensuremath{\forall}\coqdocid{x}: \coqdocid{G}, 1 \ensuremath{\times} \coqdocid{x} = \coqdocid{x}.\coqdoceol

\medskip
\noindent
\coqdockw{Lemma} \coqdocid{mulVg}: \ensuremath{\forall}\coqdocid{x}: \coqdocid{G}, \coqdocid{x}$^{-1}$ \ensuremath{\times} \coqdocid{x} = 1.\coqdoceol

\medskip
\noindent
\coqdockw{Lemma} \coqdocid{mulg\_invl}: \ensuremath{\forall}\coqdocid{x}: \coqdocid{G}, \coqdocid{cancel} (\coqdocid{mulg} \coqdocid{x}) (\coqdocid{mulg} \coqdocid{x}$^{-1}$).\coqdoceol

\medskip
\noindent
\coqdockw{Lemma} \coqdocid{mulg\_injl}: \ensuremath{\forall}\coqdocid{x}: \coqdocid{G}, \coqdocid{injective} (\coqdocid{mulg} \coqdocid{x}).\coqdoceol

\medskip
\noindent
\coqdockw{Lemma} \coqdocid{mulg1}: \ensuremath{\forall}\coqdocid{x}: \coqdocid{G}, \coqdocid{x} \ensuremath{\times} 1 = \coqdocid{x}.\coqdoceol

\medskip
\noindent
\coqdockw{Lemma} \coqdocid{invg1}: 1$^{-1}$ = 1.\coqdoceol

\medskip
\noindent
\coqdockw{Lemma} \coqdocid{mulgV}: \ensuremath{\forall}\coqdocid{x}: \coqdocid{G}, \coqdocid{x} \ensuremath{\times} \coqdocid{x}$^{-1}$ = 1.\coqdoceol

\medskip
\noindent
\coqdockw{Lemma} \coqdocid{mulg\_invr}: \ensuremath{\forall}\coqdocid{x}: \coqdocid{G}, \coqdocid{monic} (\coqdocid{mulgr} \coqdocid{x}) (\coqdocid{mulgr} \coqdocid{x}$^{-1}$).\coqdoceol

\medskip
\noindent
\coqdockw{Lemma} \coqdocid{mulg\_injr}: \ensuremath{\forall}\coqdocid{x}: \coqdocid{G}, \coqdocid{injective} (\coqdocid{mulgr} \coqdocid{x}).\coqdoceol

\medskip
\noindent
\coqdockw{Lemma} \coqdocid{invg\_inv}: \coqdocid{monic} \coqdocid{invg} \coqdocid{invg}.\coqdoceol

\medskip
\noindent
\coqdockw{Lemma} \coqdocid{invg\_inj}: \coqdocid{injective} \coqdocid{invg}.\coqdoceol

\medskip
\noindent
\coqdockw{Lemma} \coqdocid{invg\_mul}: \ensuremath{\forall}\coqdocid{x$_1$} \coqdocid{x$_2$}: \coqdocid{G}, (\coqdocid{x$_2$} \ensuremath{\times} \coqdocid{x$_1$})$^{-1}$ = \coqdocid{x$_1$}$^{-1}$ \ensuremath{\times} \coqdocid{x$_2$}$^{-1}$.\coqdoceol

\medskip
\noindent
\coqdockw{Lemma} \coqdocid{mulVg\_invl}: \ensuremath{\forall}\coqdocid{x}: \coqdocid{G}, \coqdocid{monic} (\coqdocid{mulg} \coqdocid{x}$^{-1}$) (\coqdocid{mulg} \coqdocid{x}).\coqdoceol

\medskip
\noindent
\coqdockw{Lemma} \coqdocid{mulVg\_invr}: \ensuremath{\forall}\coqdocid{x}, \coqdocid{monic} (\coqdocid{mulgr} \coqdocid{x}$^{-1}$) (\coqdocid{mulgr} \coqdocid{x}).\coqdoceol

\medskip
\noindent
\coqdockw{Theorem} \coqdocid{mulg\_s$_1$}: \ensuremath{\forall}\coqdocid{a} \coqdocid{b}:\coqdocid{G}, (\coqdocid{b} \ensuremath{\times} \coqdocid{a}$^{-1}$) \ensuremath{\times} \coqdocid{a}  = \coqdocid{b}.\coqdoceol

\medskip
\noindent
\coqdockw{Theorem} \coqdocid{mulg\_s$_2$}: \ensuremath{\forall}\coqdocid{a} \coqdocid{b}:\coqdocid{G}, (\coqdocid{b} \ensuremath{\times} \coqdocid{a}) \ensuremath{\times} \coqdocid{a}$^{-1}$ = \coqdocid{b}.\coqdoceol

\medskip
\noindent
\coqdockw{End} \coqdocid{GroupIdentities}.\coqdoceol

\medskip
\noindent
\coqdockw{Definition} \coqdocid{conjg} (\coqdocid{G}: \coqdocid{finGroup}) (\coqdocid{x} \coqdocid{y}: \coqdocid{G}):=  \coqdocid{x}$^{-1}$ \ensuremath{\times} \coqdocid{y} \ensuremath{\times} \coqdocid{x}.\coqdoceol

\medskip
\noindent
\coqdockw{Section} \coqdocid{Conjugation}.\coqdoceol

\medskip
\noindent
\coqdockw{Variable} \coqdocid{G}: \coqdocid{finGroup}.\coqdoceol

\medskip
\noindent
\coqdockw{Lemma} \coqdocid{conjgE}: \ensuremath{\forall}\coqdocid{x} \coqdocid{y}: \coqdocid{G}, \coqdocid{x}$^{\coqdocid{y}}$ = \coqdocid{y}$^{-1}$ \ensuremath{\times} \coqdocid{x} \ensuremath{\times} \coqdocid{y}.\coqdoceol

\medskip
\noindent
\coqdockw{Lemma} \coqdocid{conjg1}: \coqdocid{conjg} 1 =$_1$ \coqdocid{id}.\coqdoceol

\medskip
\noindent
\coqdockw{Lemma} \coqdocid{conj1g}: \ensuremath{\forall}\coqdocid{x}: \coqdocid{G}, 1$^{\coqdocid{x}}$ = 1.\coqdoceol

\medskip
\noindent
\coqdockw{Lemma} \coqdocid{conjg\_mul}: \ensuremath{\forall}\coqdocid{x$_1$} \coqdocid{x$_2$} \coqdocid{y}: \coqdocid{G}, (\coqdocid{x$_1$} \ensuremath{\times} \coqdocid{x$_2$})$^{\coqdocid{y}}$ = \coqdocid{x$_1$}$^{\coqdocid{y}}$ \ensuremath{\times} \coqdocid{x$_2$}$^{\coqdocid{y}}$.\coqdoceol

\medskip
\noindent
\coqdockw{Lemma} \coqdocid{conjg\_invg}: \ensuremath{\forall}\coqdocid{x} \coqdocid{y}: \coqdocid{G}, ${(\coqdocid{x}^{-1})}^{\coqdocid{y}}$ = (\coqdocid{x} $^{\coqdocid{y}}$)$^{-1}$.\coqdoceol

\medskip
\noindent
\coqdockw{Lemma} \coqdocid{conjg\_conj}: \ensuremath{\forall}\coqdocid{x} \coqdocid{y$_1$} \coqdocid{y$_2$}: \coqdocid{G}, (\coqdocid{x}$^{\coqdocid{y$_1$}}$)$^{\coqdocid{y$_2$}}$ = \coqdocid{x}$^{\coqdocid{y$_1$} \ensuremath{\times} \coqdocid{y$_2$}}$.\coqdoceol

\medskip
\noindent
\coqdockw{Lemma} \coqdocid{conjg\_inv}: \ensuremath{\forall}\coqdocid{y}: \coqdocid{G}, \coqdocid{monic} (\coqdocid{conjg} \coqdocid{y}) (\coqdocid{conjg} \coqdocid{y}$^{-1}$).\coqdoceol

\medskip
\noindent
\coqdockw{Lemma} \coqdocid{conjg\_invV}: \ensuremath{\forall}\coqdocid{y}: \coqdocid{G}, \coqdocid{monic} (\coqdocid{conjg} \coqdocid{y}$^{-1}$) (\coqdocid{conjg} \coqdocid{y}).\coqdoceol

\medskip
\noindent
\coqdockw{Lemma} \coqdocid{conjg\_inj}: \ensuremath{\forall}\coqdocid{y}: \coqdocid{G}, \coqdocid{injective} (\coqdocid{conjg} \coqdocid{y}).\coqdoceol

\medskip
\noindent
\coqdockw{Definition} \coqdocid{conjg\_fp} (\coqdocid{y} \coqdocid{x}: \coqdocid{G}):= \coqdocid{x}$^{\coqdocid{y}}$ =${_d}$ \coqdocid{x}.\coqdoceol

\medskip
\noindent
\coqdockw{Definition} \coqdocid{commg} (\coqdocid{x} \coqdocid{y}: \coqdocid{G}):= \coqdocid{x} \ensuremath{\times} \coqdocid{y} = \coqdocid{y} \ensuremath{\times} \coqdocid{x}.\coqdoceol

\medskip
\noindent
\coqdockw{Lemma} \coqdocid{conjg\_fpP}: \ensuremath{\forall}\coqdocid{x} \coqdocid{y}: \coqdocid{G}, \coqdocid{reflect} (\coqdocid{commg} \coqdocid{x} \coqdocid{y}) (\coqdocid{conjg\_fp} \coqdocid{y} \coqdocid{x}).\coqdoceol

\medskip
\noindent
\coqdockw{Lemma} \coqdocid{conjg\_fp\_sym}: \ensuremath{\forall}\coqdocid{x} \coqdocid{y}: \coqdocid{G}, \coqdocid{conjg\_fp} \coqdocid{x} \coqdocid{y} = \coqdocid{conjg\_fp} \coqdocid{y} \coqdocid{x}.\coqdoceol

\medskip
\noindent
\coqdockw{End} \coqdocid{Conjugation}.\coqdoceol

\medskip
\noindent
\coqdockw{Section} \coqdocid{SubGroup}.\coqdoceol

\medskip
\noindent
\coqdockw{Variables} (\coqdocid{G}: \coqdocid{finGroup}) (\coqdocid{H}: \coqdocid{set} \coqdocid{G}).\coqdoceol

\medskip
\noindent
\coqdockw{Definition} \coqdocid{lcoset} \coqdocid{x}: \coqdocid{set} \coqdocid{G}:= \coqdocid{fun} \coqdocid{y} \ensuremath{\Rightarrow} \coqdocid{H} (\coqdocid{x}$^{-1}$ \ensuremath{\times} \coqdocid{y}).\coqdoceol
\noindent
\coqdockw{Definition} \coqdocid{rcoset} \coqdocid{x}: \coqdocid{set} \coqdocid{G}:= \coqdocid{fun} \coqdocid{y} \ensuremath{\Rightarrow} \coqdocid{H} (\coqdocid{y} \ensuremath{\times} \coqdocid{x}$^{-1}$).\coqdoceol

\medskip
\noindent
\coqdockw{Definition} \coqdocid{subgrpb}:= \coqdocid{H} 1 \&\& \coqdocid{subset} \coqdocid{H} (\coqdocid{fun} \coqdocid{x} \ensuremath{\Rightarrow} \coqdocid{subset} \coqdocid{H} (\coqdocid{rcoset} \coqdocid{x})).\coqdoceol

\medskip
\noindent
\coqdockw{Definition} \coqdocid{subgrp}: \coqdocid{Prop}:= \coqdocid{subgrpb}.\coqdoceol

\medskip
\noindent
\coqdockw{Lemma} \coqdocid{subgrpP}: \coqdocid{reflect} (\coqdocid{H} 1 \ensuremath{\land} \ensuremath{\forall}\coqdocid{x} \coqdocid{y}, \coqdocid{H} \coqdocid{x} \ensuremath{\rightarrow} \coqdocid{H} \coqdocid{y} \ensuremath{\rightarrow} \coqdocid{rcoset} \coqdocid{x} \coqdocid{y}) \coqdocid{subgrpb}.\coqdoceol

\medskip
\noindent
\coqdockw{Hypothesis} \coqdocid{Hh}: \coqdocid{subgrp}.\coqdoceol

\medskip
\noindent
\coqdockw{Lemma} \coqdocid{subgrp1}: \coqdocid{H} 1.\coqdoceol

\medskip
\noindent
\coqdockw{Lemma} \coqdocid{subgrpV}: \ensuremath{\forall}\coqdocid{x}, \coqdocid{H} \coqdocid{x} \ensuremath{\rightarrow} \coqdocid{H} \coqdocid{x}$^{-1}$.\coqdoceol

\medskip
\noindent
\coqdockw{Lemma} \coqdocid{subgrpM}: \ensuremath{\forall}\coqdocid{x} \coqdocid{y}, \coqdocid{H} \coqdocid{x} \ensuremath{\rightarrow} \coqdocid{H} \coqdocid{y} \ensuremath{\rightarrow} \coqdocid{H} (\coqdocid{x} \ensuremath{\times} \coqdocid{y}).\coqdoceol

\medskip
\noindent
\coqdockw{Lemma} \coqdocid{subgrpMl}: \ensuremath{\forall}\coqdocid{x} \coqdocid{y}, \coqdocid{H} \coqdocid{x} \ensuremath{\rightarrow} \coqdocid{H} (\coqdocid{x} \ensuremath{\times} \coqdocid{y}) = \coqdocid{H} \coqdocid{y}.\coqdoceol

\medskip
\noindent
\coqdockw{Lemma} \coqdocid{subgrpMr}: \ensuremath{\forall}\coqdocid{x} \coqdocid{y}, \coqdocid{H} \coqdocid{x} \ensuremath{\rightarrow} \coqdocid{H} (\coqdocid{y} \ensuremath{\times} \coqdocid{x}) = \coqdocid{H} \coqdocid{y}.\coqdoceol

\medskip
\noindent
\coqdockw{Lemma} \coqdocid{subgrpVl}: \ensuremath{\forall}\coqdocid{x},  \coqdocid{H} \coqdocid{x}$^{-1}$ \ensuremath{\rightarrow}  \coqdocid{H} \coqdocid{x}.\coqdoceol

\medskip
\noindent
\coqdockw{Definition} \coqdocid{subFinGroup}: \coqdocid{finGroup}.\coqdoceol

\medskip
\noindent
\coqdockw{End} \coqdocid{SubGroup}.\coqdoceol

\medskip
\noindent
\coqdockw{Lemma} \coqdocid{subgrp\_of\_group}: \ensuremath{\forall}\coqdocid{G}: \coqdocid{finGroup}, \coqdocid{subgrp} \coqdocid{G}.\coqdoceol

\medskip
\noindent
\coqdocid{Coercion} \coqdocid{subgrp\_of\_group}: \coqdocid{finGroup} >-> \coqdocid{subgrp}.\coqdoceol

\medskip
\noindent
\coqdockw{Section} \coqdocid{LaGrange}.\coqdoceol

\medskip
\noindent
\coqdockw{Variables} (\coqdocid{G}: \coqdocid{finGroup}) (\coqdocid{H}: \coqdocid{set} \coqdocid{G}).\coqdoceol
\noindent
\coqdockw{Hypothesis} (\coqdocid{Hh}: \coqdocid{subgrp} \coqdocid{H}).\coqdoceol

\medskip
\noindent
\coqdockw{Lemma} \coqdocid{rcoset\_refl}: \ensuremath{\forall}\coqdocid{x}, \coqdocid{rcoset} \coqdocid{H} \coqdocid{x} \coqdocid{x}.\coqdoceol

\medskip
\noindent
\coqdockw{Lemma} \coqdocid{rcoset\_sym}: \ensuremath{\forall}\coqdocid{x} \coqdocid{y}, \coqdocid{rcoset} \coqdocid{H} \coqdocid{x} \coqdocid{y} = \coqdocid{rcoset} \coqdocid{H} \coqdocid{y} \coqdocid{x}.\coqdoceol

\medskip
\noindent
\coqdockw{Lemma} \coqdocid{rcoset\_trans}: \ensuremath{\forall}\coqdocid{x} \coqdocid{y}, \coqdocid{connect} (\coqdocid{rcoset} \coqdocid{H}) \coqdocid{x} \coqdocid{y} = \coqdocid{rcoset} \coqdocid{H} \coqdocid{x} \coqdocid{y}.\coqdoceol

\medskip
\noindent
\coqdockw{Lemma} \coqdocid{rcoset\_csym}: \coqdocid{connect\_sym} (\coqdocid{rcoset} \coqdocid{H}).\coqdoceol

\medskip
\noindent
\coqdockw{Lemma} \coqdocid{rcoset1}: \coqdocid{rcoset} \coqdocid{H} 1 =$_1$ \coqdocid{H}.\coqdoceol

\medskip
\noindent
\coqdockw{Lemma} \coqdocid{card\_rcoset}: \ensuremath{\forall}\coqdocid{x}, \coqdocid{card} (\coqdocid{rcoset} \coqdocid{H} \coqdocid{x}) = \coqdocid{card} \coqdocid{H}.\coqdoceol

\medskip
\noindent
\coqdockw{Definition} \coqdocid{rindex}:= \coqdocid{n\_comp} (\coqdocid{rcoset} \coqdocid{H}).\coqdoceol

\medskip
\noindent
\coqdockw{Theorem} \coqdocid{rLaGrange}: \ensuremath{\forall}\coqdocid{K}: \coqdocid{set} \coqdocid{G},\coqdoceol
\coqdocindent{1.00em}
\coqdocid{subgrp} \coqdocid{K} \ensuremath{\rightarrow} \coqdocid{subset} \coqdocid{H} \coqdocid{K} \ensuremath{\rightarrow} \coqdocid{card} \coqdocid{H} \ensuremath{\times} \coqdocid{rindex} \coqdocid{K} = \coqdocid{card} \coqdocid{K}.\coqdoceol

\medskip
\noindent
\coqdockw{Theorem} \coqdocid{sugrp\_divn}: \ensuremath{\forall}\coqdocid{K}: \coqdocid{set} \coqdocid{G},\coqdoceol
\coqdocindent{1.00em}
\coqdocid{subgrp} \coqdocid{K} \ensuremath{\rightarrow} \coqdocid{subset} \coqdocid{H} \coqdocid{K} \ensuremath{\rightarrow}  \coqdocid{card} \coqdocid{H} \ensuremath{|}  \coqdocid{card} \coqdocid{K}.\coqdoceol

\medskip
\noindent
\coqdockw{Lemma} \coqdocid{lcoset\_refl}: \ensuremath{\forall}\coqdocid{x}, \coqdocid{lcoset} \coqdocid{H} \coqdocid{x} \coqdocid{x}.\coqdoceol

\medskip
\noindent
\coqdockw{Lemma} \coqdocid{lcoset\_sym}: \ensuremath{\forall}\coqdocid{x} \coqdocid{y}, \coqdocid{lcoset} \coqdocid{H} \coqdocid{x} \coqdocid{y} = \coqdocid{lcoset} \coqdocid{H} \coqdocid{y} \coqdocid{x}.\coqdoceol

\medskip
\noindent
\coqdockw{Lemma} \coqdocid{lcoset\_trans}: \ensuremath{\forall}\coqdocid{x} \coqdocid{y}, \coqdocid{connect} (\coqdocid{lcoset} \coqdocid{H}) \coqdocid{x} \coqdocid{y} = \coqdocid{lcoset} \coqdocid{H} \coqdocid{x} \coqdocid{y}.\coqdoceol

\medskip
\noindent
\coqdockw{Lemma} \coqdocid{lcoset\_csym}: \coqdocid{connect\_sym} (\coqdocid{lcoset} \coqdocid{H}).\coqdoceol

\medskip
\noindent
\coqdockw{Lemma} \coqdocid{lcoset1}: \coqdocid{lcoset} \coqdocid{H} 1 =$_1$ \coqdocid{H}.\coqdoceol

\medskip
\noindent
\coqdockw{Lemma} \coqdocid{card\_lcoset}: \ensuremath{\forall}\coqdocid{x}, \coqdocid{card} (\coqdocid{lcoset} \coqdocid{H} \coqdocid{x}) = \coqdocid{card} \coqdocid{H}.\coqdoceol

\medskip
\noindent
\coqdockw{Definition} \coqdocid{lindex}:= \coqdocid{n\_comp} (\coqdocid{lcoset} \coqdocid{H}).\coqdoceol

\medskip
\noindent
\coqdockw{Theorem} \coqdocid{lLaGrange}: \ensuremath{\forall}\coqdocid{K}: \coqdocid{set} \coqdocid{G},\coqdoceol
\coqdocindent{1.00em}
\coqdocid{subgrp} \coqdocid{K} \ensuremath{\rightarrow} \coqdocid{subset} \coqdocid{H} \coqdocid{K} \ensuremath{\rightarrow} \coqdocid{card} \coqdocid{H} \ensuremath{\times} \coqdocid{lindex} \coqdocid{K} = \coqdocid{card} \coqdocid{K}.\coqdoceol

\medskip
\noindent
\coqdockw{End} \coqdocid{LaGrange}.\coqdoceol

\medskip
\noindent
\coqdockw{Section} \coqdocid{FinPart}.\coqdoceol

\medskip
\noindent
\coqdockw{Variables} (\coqdocid{G}: \coqdocid{finGroup}) (\coqdocid{H}: \coqdocid{set} \coqdocid{G}) (\coqdocid{a}: \coqdocid{G}).\coqdoceol
\noindent
\coqdockw{Hypothesis} \coqdocid{Ha}: \coqdocid{H} \coqdocid{a}.\coqdoceol
\noindent
\coqdockw{Hypothesis} \coqdocid{Hstable}: \ensuremath{\forall}\coqdocid{x} \coqdocid{y}, \coqdocid{H} \coqdocid{x} \ensuremath{\rightarrow} \coqdocid{H} \coqdocid{y} \ensuremath{\rightarrow} \coqdocid{H} (\coqdocid{x} \ensuremath{\times} \coqdocid{y}).\coqdoceol

\medskip
\noindent
\coqdockw{Lemma} \coqdocid{heqah}: (\coqdocid{lcoset} \coqdocid{H} \coqdocid{a}) =$_1$ \coqdocid{H}.\coqdoceol

\medskip
\noindent
\coqdockw{Lemma} \coqdocid{heqxh}: \ensuremath{\forall}\coqdocid{x}, \coqdocid{H} \coqdocid{x} \ensuremath{\rightarrow} (\coqdocid{lcoset} \coqdocid{H} \coqdocid{x}) =$_1$ \coqdocid{H}.\coqdoceol

\medskip
\noindent
\coqdockw{Lemma} \coqdocid{heqhx}: \ensuremath{\forall}\coqdocid{x}, \coqdocid{H} \coqdocid{x}\ensuremath{\rightarrow} (\coqdocid{rcoset} \coqdocid{H} \coqdocid{x}) =$_1$ \coqdocid{H}.\coqdoceol

\medskip
\noindent
\coqdockw{Lemma} \coqdocid{finstbl\_sbgrp1}: \coqdocid{H} 1.\coqdoceol

\medskip
\noindent
\coqdockw{Lemma} \coqdocid{finstbl\_mulV}: \ensuremath{\forall}\coqdocid{x}, \coqdocid{H} \coqdocid{x} \ensuremath{\rightarrow} \coqdocid{H} \coqdocid{x}$^{-1}$.\coqdoceol

\medskip
\noindent
\coqdockw{Lemma} \coqdocid{finstbl\_sbgrp}: \coqdocid{subgrp} \coqdocid{H}.\coqdoceol

\medskip
\noindent
\coqdockw{End} \coqdocid{FinPart}.\coqdoceol

\medskip
\noindent
\coqdockw{Section} \coqdocid{Eq}.\coqdoceol

\medskip
\noindent
\coqdockw{Variable} \coqdocid{G}: \coqdocid{finGroup}.\coqdoceol

\medskip
\noindent
\coqdockw{Theorem} \coqdocid{eq\_subgroup}: \ensuremath{\forall}\coqdocid{a} \coqdocid{b}: \coqdocid{set} \coqdocid{G}, \coqdocid{a} =$_1$ \coqdocid{b} \ensuremath{\rightarrow} \coqdocid{subgrpb} \coqdocid{a} = \coqdocid{subgrpb} \coqdocid{b}.\coqdoceol

\medskip
\noindent
\coqdockw{End} \coqdocid{Eq}.\coqdoceol

\medskip
\noindent
\coqdockw{Section} \coqdocid{SubProd}.\coqdoceol
\noindent
\noindent
\coqdockw{Variable} \coqdocid{G}: \coqdocid{finGroup}.\coqdoceol

\medskip
\noindent
\coqdockw{Section} \coqdocid{SubProd\_subgrp}.\coqdoceol

\medskip
\noindent
\coqdockw{Variables} (\coqdocid{H} \coqdocid{K}: \coqdocid{set} \coqdocid{G}).\coqdoceol
\noindent
\coqdockw{Hypothesis} \coqdocid{h\_subgroup}: \coqdocid{subgrp} \coqdocid{H}.\coqdoceol
\noindent
\coqdockw{Hypothesis} \coqdocid{k\_subgroup}: \coqdocid{subgrp} \coqdocid{K}.\coqdoceol

\medskip
\noindent
\coqdockw{Lemma} \coqdocid{subprod\_sbgrp}: \coqdocid{prod} \coqdocid{H} \coqdocid{K} =$_1$ \coqdocid{prod} \coqdocid{K} \coqdocid{H} \ensuremath{\rightarrow} \coqdocid{subgrp} (\coqdocid{prod} \coqdocid{H} \coqdocid{K}).\coqdoceol

\medskip
\noindent
\coqdockw{Lemma} \coqdocid{sbgrp\_subprod}: \coqdocid{subgrp} (\coqdocid{prod} \coqdocid{H} \coqdocid{K}) \ensuremath{\rightarrow} \coqdocid{prod} \coqdocid{H} \coqdocid{K} =$_1$ \coqdocid{prod} \coqdocid{K} \coqdocid{H}.\coqdoceol

\medskip
\noindent
\coqdockw{End} \coqdocid{SubProd\_subgrp}.\coqdoceol

\medskip
\noindent
\coqdockw{Variables} (\coqdocid{H} \coqdocid{K}: \coqdocid{set} \coqdocid{G}).\coqdoceol
\noindent
\coqdockw{Hypothesis} \coqdocid{h\_subgroup}: \coqdocid{subgrp} \coqdocid{H}.\coqdoceol
\noindent
\coqdockw{Hypothesis} \coqdocid{k\_subgroup}: \coqdocid{subgrp} \coqdocid{K}.\coqdoceol

\medskip
\noindent
\coqdockw{Lemma} \coqdocid{sbgrphk\_sbgrpkh}: \coqdocid{subgrpb} (\coqdocid{prod} \coqdocid{H} \coqdocid{K}) =  \coqdocid{subgrpb} (\coqdocid{prod} \coqdocid{K} \coqdocid{H}).\coqdoceol

\medskip
\noindent
\coqdockw{End} \coqdocid{SubProd}.\coqdoceol

\coqdocmodule{action}

\medskip
\noindent
\coqdockw{Section} \coqdocid{Action}.\coqdoceol

\medskip
\noindent
\coqdockw{Variable} (\coqdocid{G}: \coqdocid{finGroup}) (\coqdocid{H}: \coqdocid{set} \coqdocid{G}).\coqdoceol

\medskip
\noindent
\coqdockw{Hypothesis} \coqdocid{sgrp\_h}: \coqdocid{subgrp} \coqdocid{H}.\coqdoceol
\noindent
\coqdockw{Variable} \coqdocid{s}: \coqdocid{finType}.\coqdoceol

\medskip
\noindent
\coqdockw{Variable} \coqdocid{to}: \coqdocid{G} \ensuremath{\rightarrow} (\coqdocid{s} \ensuremath{\rightarrow} \coqdocid{s}).\coqdoceol
\noindent
\coqdockw{Hypothesis} \coqdocid{to\_bij}: \ensuremath{\forall}\coqdocid{x}, \coqdocid{H} \coqdocid{x} \ensuremath{\rightarrow} \coqdocid{bijective} (\coqdocid{to} \coqdocid{x}).\coqdoceol
\noindent
\coqdockw{Hypothesis} \coqdocid{to\_morph}: \ensuremath{\forall}(\coqdocid{x} \coqdocid{y}: \coqdocid{G}) \coqdocid{z}, \coqdoceol
\coqdocindent{1.00em}
\coqdocid{H} \coqdocid{x} \ensuremath{\rightarrow} \coqdocid{H} \coqdocid{y} \ensuremath{\rightarrow} \coqdocid{to} (\coqdocid{x} \ensuremath{\times} \coqdocid{y}) \coqdocid{z} = \coqdocid{to} \coqdocid{x} (\coqdocid{to} \coqdocid{y} \coqdocid{z}).\coqdoceol

\medskip
\noindent
\coqdockw{Theorem} \coqdocid{to\_1}: \ensuremath{\forall}\coqdocid{x}, \coqdocid{to} 1 \coqdocid{x} = \coqdocid{x}.\coqdoceol

\medskip
\noindent
\coqdockw{Definition} \coqdocid{stabiliser} \coqdocid{a}:= \coqdocid{setI} (\coqdocid{fun} \coqdocid{x} \ensuremath{\Rightarrow} ((\coqdocid{to} \coqdocid{x} \coqdocid{a}) =${_d}$ \coqdocid{a})) \coqdocid{H}.\coqdoceol
\noindent
\coqdockw{Definition} \coqdocid{orbit} \coqdocid{a}:= \coqdocid{image} (\coqdocid{fun} \coqdocid{z} \ensuremath{\Rightarrow} \coqdocid{to} \coqdocid{z} \coqdocid{a}) \coqdocid{H}.\coqdoceol

\medskip
\noindent
\coqdockw{Theorem} \coqdocid{orbit\_to}: \ensuremath{\forall}\coqdocid{a} \coqdocid{x}, \coqdocid{H} \coqdocid{x} \ensuremath{\rightarrow} \coqdocid{orbit} \coqdocid{a} (\coqdocid{to} \coqdocid{x} \coqdocid{a}).\coqdoceol

\medskip
\noindent
\coqdockw{Lemma} \coqdocid{orbit\_refl}: \ensuremath{\forall}\coqdocid{x}, \coqdocid{orbit} \coqdocid{x} \coqdocid{x}.\coqdoceol

\medskip
\noindent
\coqdockw{Lemma} \coqdocid{orbit\_sym}: \ensuremath{\forall}\coqdocid{x} \coqdocid{y}, \coqdocid{orbit} \coqdocid{x} \coqdocid{y} = \coqdocid{orbit} \coqdocid{y} \coqdocid{x}.\coqdoceol

\medskip
\noindent
\coqdockw{Lemma} \coqdocid{orbit\_trans}: \ensuremath{\forall}\coqdocid{x} \coqdocid{y}, \coqdocid{connect} \coqdocid{orbit} \coqdocid{x} \coqdocid{y} = \coqdocid{orbit} \coqdocid{x} \coqdocid{y}.\coqdoceol

\medskip
\noindent
\coqdockw{Lemma} \coqdocid{orbit\_csym}: \coqdocid{connect\_sym} \coqdocid{orbit}.\coqdoceol

\medskip
\noindent
\coqdockw{Definition} \coqdocid{S$_0$} \coqdocid{a}:= \coqdocid{subset} \coqdocid{H} (\coqdocid{stabiliser} \coqdocid{a}).\coqdoceol

\medskip
\noindent
\coqdockw{Theorem} \coqdocid{S0P}: \ensuremath{\forall}\coqdocid{a}, \coqdocid{reflect} (\coqdocid{orbit} \coqdocid{a} =$_1$ \coqdocid{set1} \coqdocid{a}) (\coqdocid{S$_0$} \coqdocid{a}).\coqdoceol

\medskip
\noindent
\coqdockw{Theorem} \coqdocid{stab\_1}: \ensuremath{\forall}\coqdocid{a}, \coqdocid{stabiliser} \coqdocid{a} 1.\coqdoceol

\medskip
\noindent
\coqdockw{Theorem} \coqdocid{subgr\_stab}: \ensuremath{\forall}\coqdocid{a}, \coqdocid{subgrp} (\coqdocid{stabiliser} \coqdocid{a}).\coqdoceol

\medskip
\noindent
\coqdockw{Theorem} \coqdocid{subset\_stab}: \ensuremath{\forall}\coqdocid{a}, \coqdocid{subset} (\coqdocid{stabiliser} \coqdocid{a}) \coqdocid{H}.\coqdoceol

\medskip
\noindent
\coqdockw{Theorem} \coqdocid{orbit\_from}: \ensuremath{\forall}\coqdocid{a} \coqdocid{x} (\coqdocid{Hx}: \coqdocid{orbit} \coqdocid{a} \coqdocid{x}),\coqdoceol
\coqdocindent{1.00em}
(\coqdocid{setI} (\coqdocid{roots} (\coqdocid{lcoset} (\coqdocid{stabiliser} \coqdocid{a})))  \coqdocid{H})
(\coqdocid{root} (\coqdocid{lcoset} (\coqdocid{iinv1} \coqdocid{Hx})).\coqdoceol

\medskip
\noindent
\coqdockw{Theorem} \coqdocid{card\_orbit}: \ensuremath{\forall}\coqdocid{a}, \coqdocid{card} (\coqdocid{orbit} \coqdocid{a}) = \coqdocid{lindex} (\coqdocid{stabiliser} \coqdocid{a}) \coqdocid{H}.\coqdoceol

\medskip
\noindent
\coqdockw{Theorem} \coqdocid{card\_orbit\_div}: \ensuremath{\forall}\coqdocid{a}, \coqdocid{card} (\coqdocid{orbit} \coqdocid{a}) \ensuremath{|} \coqdocid{card} \coqdocid{H}.\coqdoceol

\medskip
\noindent
\coqdockw{Variable} \coqdocid{n} \coqdocid{p}: \coqdocid{nat}.\coqdoceol
\noindent
\coqdockw{Hypothesis} \coqdocid{prime\_p}: \coqdocid{prime} \coqdocid{p}.\coqdoceol
\noindent
\coqdockw{Hypothesis} \coqdocid{card\_h}: \coqdocid{card} \coqdocid{H} = \coqdocid{p}$^{\coqdocid{n}}$.\coqdoceol

\medskip
\noindent
\coqdockw{Theorem} \coqdocid{mpl}: (\coqdocid{card} \coqdocid{s}) \% \coqdocid{p} = (\coqdocid{card} \coqdocid{S$_0$}) \% \coqdocid{p}.\coqdoceol

\medskip
\noindent
\coqdockw{End} \coqdocid{Action}.\coqdoceol

\coqdocmodule{cyclic}

\medskip
\noindent
\coqdockw{Section} \coqdocid{Phi}.\coqdoceol
\noindent
\coqdockw{Definition} \coqdocid{phi} \coqdocid{n}:= 
\coqdocid{if} \coqdocid{n} \coqdocid{is} \coqdocid{n$_1$} \ensuremath{+} 1 \coqdocid{then}
\coqdocid{card} (\coqdocid{fun} \coqdocid{x} \ensuremath{\Rightarrow} \coqdocid{coprime} \coqdocid{n} (\coqdocid{val} \coqdocid{x}))
\coqdocid{else} 0.\coqdoceol

\medskip
\noindent
\coqdockw{Theorem} \coqdocid{phi\_mult}: \ensuremath{\forall}\coqdocid{m} \coqdocid{n}, \coqdocid{coprime} \coqdocid{m} \coqdocid{n} \ensuremath{\rightarrow} \coqdocid{phi} (\coqdocid{m} \ensuremath{\times} \coqdocid{n}) = \coqdocid{phi} \coqdocid{m} \ensuremath{\times} \coqdocid{phi} \coqdocid{n}.\coqdoceol

\medskip
\noindent
\coqdockw{Theorem} \coqdocid{phi\_prime\_k}: \ensuremath{\forall}\coqdocid{p} \coqdocid{k}, \coqdocid{prime} \coqdocid{p} \ensuremath{\rightarrow}
\coqdocid{phi} \coqdocid{p}$^{\coqdocid{k} \ensuremath{+} 1}$ = \coqdocid{p}$^{\coqdocid{k} \ensuremath{+} 1}$ - \coqdocid{p}$^{\coqdocid{k}}$.\coqdoceol

\medskip
\noindent
\coqdockw{End} \coqdocid{Phi}.\coqdoceol

\medskip
\noindent
\coqdockw{Section} \coqdocid{Cyclic}.\coqdoceol

\medskip
\noindent
\coqdockw{Variable} \coqdocid{G}: \coqdocid{finGroup}.\coqdoceol

\medskip
\noindent
\coqdockw{Fixpoint} \coqdocid{gexpn} (\coqdocid{a}:\coqdocid{G}) (\coqdocid{n}: \coqdocid{nat}) \{\coqdocid{struct} \coqdocid{n}\}: \coqdocid{G}:=\coqdoceol
\coqdocindent{1.00em}
\coqdocid{if} \coqdocid{n} \coqdocid{is} \coqdocid{n$_1$} \ensuremath{+} 1 \coqdocid{then} \coqdocid{a} \ensuremath{\times} (\coqdocid{gexpn} \coqdocid{a} \coqdocid{n$_1$}) \coqdocid{else} 1.\coqdoceol

\medskip
\noindent
\coqdockw{Theorem} \coqdocid{gexpn0}: \ensuremath{\forall}\coqdocid{a}, \coqdocid{gexpn} \coqdocid{a} 0 = 1.\coqdoceol

\medskip
\noindent
\coqdockw{Theorem} \coqdocid{gexpn1}: \ensuremath{\forall}\coqdocid{a}, \coqdocid{gexpn} \coqdocid{a} 1 = \coqdocid{a}.\coqdoceol

\medskip
\noindent
\coqdockw{Theorem} \coqdocid{gexp1n}: \ensuremath{\forall}\coqdocid{n}, \coqdocid{gexpn} 1 \coqdocid{n} = 1.\coqdoceol

\medskip
\noindent
\coqdockw{Theorem} \coqdocid{gexpnS}: \ensuremath{\forall}\coqdocid{a} \coqdocid{n}, \coqdocid{gexpn} \coqdocid{a} (\coqdocid{n} \ensuremath{+} 1)) = \coqdocid{a} \ensuremath{\times} \coqdocid{gexpn} \coqdocid{a} \coqdocid{n}.\coqdoceol

\medskip
\noindent
\coqdockw{Theorem} \coqdocid{gexpn\_h}: \ensuremath{\forall}\coqdocid{n} \coqdocid{a} \coqdocid{H}, \coqdocid{subgrp} \coqdocid{H} \ensuremath{\rightarrow} \coqdocid{H} \coqdocid{a} \ensuremath{\rightarrow} \coqdocid{H} (\coqdocid{gexpn} \coqdocid{a} \coqdocid{n}).\coqdoceol

\medskip
\noindent
\coqdockw{Theorem} \coqdocid{gexpn\_add}: \ensuremath{\forall}\coqdocid{a} \coqdocid{n} \coqdocid{m}, \coqdocid{gexpn} \coqdocid{a} \coqdocid{n} \ensuremath{\times} \coqdocid{gexpn} \coqdocid{a} \coqdocid{m} = \coqdocid{gexpn} \coqdocid{a} (\coqdocid{n} + \coqdocid{m}).\coqdoceol

\medskip
\noindent
\coqdockw{Theorem} \coqdocid{gexpn\_mul}: \ensuremath{\forall}\coqdocid{a} \coqdocid{n} \coqdocid{m}, \coqdocid{gexpn} (\coqdocid{gexpn} \coqdocid{a} \coqdocid{n}) \coqdocid{m} = \coqdocid{gexpn} \coqdocid{a} (\coqdocid{n} \ensuremath{\times} \coqdocid{m}).\coqdoceol

\medskip
\noindent
\coqdockw{Fixpoint} \coqdocid{seq\_fn} (\coqdocid{f}: \coqdocid{G} \ensuremath{\rightarrow} \coqdocid{G}) (\coqdocid{n}: \coqdocid{nat}) (\coqdocid{a}: \coqdocid{G}) (\coqdocid{L}: \coqdocid{seq} \coqdocid{G}) \{\coqdocid{struct} \coqdocid{n}\}: \coqdocid{seq} \coqdocid{G}:=\coqdoceol
\coqdocindent{1.00em}
\coqdocid{if} \coqdocid{n} \coqdocid{is} \coqdocid{n$_1$} \ensuremath{+} 1 \coqdocid{then}\coqdoceol
\coqdocindent{2.50em}
\coqdocid{if} \coqdocid{negb} (\coqdocid{L} \coqdocid{a}) \coqdocid{then} \coqdocid{seq\_fn} \coqdocid{f} \coqdocid{n$_1$} (\coqdocid{f} \coqdocid{a}) (\coqdocid{Adds} \coqdocid{a} \coqdocid{L}) \coqdocid{else} \coqdocid{L} \coqdocid{else} \coqdocid{L}.\coqdoceol

\medskip
\noindent
\coqdockw{Definition} \coqdocid{seq\_f} \coqdocid{f} \coqdocid{a}:= \coqdocid{seq\_fn} \coqdocid{f} (\coqdocid{card} \coqdocid{G}) \coqdocid{a} (\coqdocid{Seq0} \coqdocid{\_}).\coqdoceol

\medskip
\noindent
\coqdockw{Definition} \coqdocid{cyclic} \coqdocid{a}:= \coqdocid{seq\_f} (\coqdocid{fun} \coqdocid{x} \ensuremath{\Rightarrow} \coqdocid{a} \ensuremath{\times} \coqdocid{x}) 1.\coqdoceol

\medskip
\noindent
\coqdockw{Theorem} \coqdocid{cyclic1}: \ensuremath{\forall}\coqdocid{a}, \coqdocid{cyclic} \coqdocid{a} 1.\coqdoceol

\medskip
\noindent
\coqdockw{Theorem} \coqdocid{cyclicP}: \ensuremath{\forall}\coqdocid{a} \coqdocid{b}, \coqdocid{reflect} (\ensuremath{\exists} \coqdocid{n}, \coqdocid{gexpn} \coqdocid{a} \coqdocid{n} =${_d}$ \coqdocid{b}) (\coqdocid{cyclic} \coqdocid{a} \coqdocid{b}).\coqdoceol

\medskip
\noindent
\coqdockw{Theorem} \coqdocid{cyclic\_h}: \ensuremath{\forall}\coqdocid{a} \coqdocid{H}, \coqdocid{subgrp} \coqdocid{H} \ensuremath{\rightarrow} \coqdocid{H} \coqdocid{a} \ensuremath{\rightarrow} \coqdocid{subset} (\coqdocid{cyclic} \coqdocid{a}) \coqdocid{H}.\coqdoceol

\medskip
\noindent
\coqdockw{Theorem} \coqdocid{cyclic\_min}: \ensuremath{\forall}\coqdocid{a} \coqdocid{b}, \coqdoceol
\coqdocindent{1.00em}
\coqdocid{cyclic} \coqdocid{a} \coqdocid{b} \ensuremath{\rightarrow} \ensuremath{\exists} \coqdocid{m}, (\coqdocid{m} < \coqdocid{card} (\coqdocid{cyclic} \coqdocid{a})) \&\& (\coqdocid{gexpn} \coqdocid{a} \coqdocid{m} =${_d}$ \coqdocid{b}).\coqdoceol

\medskip
\noindent
\coqdockw{Theorem} \coqdocid{cyclic\_in}: \ensuremath{\forall}\coqdocid{a} \coqdocid{m}, \coqdocid{cyclic} \coqdocid{a} (\coqdocid{gexpn} \coqdocid{a} \coqdocid{m}).\coqdoceol

\medskip
\noindent
\coqdockw{Theorem} \coqdocid{subgr\_cyclic}: \ensuremath{\forall}\coqdocid{a}, \coqdocid{subgrp} (\coqdocid{cyclic} \coqdocid{a}).\coqdoceol

\medskip
\noindent
\coqdockw{Theorem} \coqdocid{cyclic\_expn\_card}: \ensuremath{\forall}\coqdocid{a}, \coqdocid{gexpn} \coqdocid{a} (\coqdocid{card} (\coqdocid{cyclic} \coqdocid{a})) =${_d}$ 1.\coqdoceol

\medskip
\noindent
\coqdockw{Theorem} \coqdocid{cyclic\_div\_card}: \ensuremath{\forall}\coqdocid{a} \coqdocid{n}, \coqdocid{card} (\coqdocid{cyclic} \coqdocid{a}) \ensuremath{|} \coqdocid{n}) = (\coqdocid{gexpn} \coqdocid{a} \coqdocid{n} =${_d}$ 1).\coqdoceol

\medskip
\noindent
\coqdockw{Theorem} \coqdocid{cyclic\_div\_g}:
\ensuremath{\forall}\coqdocid{a},  \coqdocid{card} (\coqdocid{cyclic} \coqdocid{a}) \ensuremath{|} \coqdocid{card} \coqdocid{G}.\coqdoceol

\coqdocmodule{normal}

\medskip
\noindent
\coqdockw{Section} \coqdocid{Normal}.\coqdoceol

\medskip
\noindent
\coqdockw{Variables} (\coqdocid{G}: \coqdocid{finGroup}) (\coqdocid{H} \coqdocid{K}: \coqdocid{set} \coqdocid{G}).\coqdoceol

\medskip
\noindent
\coqdockw{Hypothesis} \coqdocid{sgrp\_h}: \coqdocid{subgrp} \coqdocid{H}.\coqdoceol
\noindent
\coqdockw{Hypothesis} \coqdocid{sgrp\_k}: \coqdocid{subgrp} \coqdocid{K}.\coqdoceol
\noindent
\coqdockw{Hypothesis} \coqdocid{subset\_hk}: \coqdocid{subset} \coqdocid{H} \coqdocid{K}.\coqdoceol

\medskip
\noindent
\coqdockw{Definition} \coqdocid{conjsg} \coqdocid{x} \coqdocid{y}:= \coqdocid{H}(\coqdocid{y}$^{\coqdocid{x}}$).\coqdoceol

\medskip
\noindent
\coqdockw{Theorem} \coqdocid{conjsg1}: \ensuremath{\forall}\coqdocid{x}, \coqdocid{conjsg} \coqdocid{x} 1.\coqdoceol

\medskip
\noindent
\coqdockw{Theorem} \coqdocid{conjs$_1$g}: \ensuremath{\forall}\coqdocid{x}, \coqdocid{conjsg} 1 \coqdocid{x} = \coqdocid{H} \coqdocid{x}.\coqdoceol

\medskip
\noindent
\coqdockw{Theorem} \coqdocid{conjsg\_inv}: \ensuremath{\forall}\coqdocid{x} \coqdocid{y}, \coqdocid{conjsg} \coqdocid{x} \coqdocid{y} \ensuremath{\rightarrow} \coqdocid{conjsg} \coqdocid{x} \coqdocid{y}$^{-1}$.\coqdoceol

\medskip
\noindent
\coqdockw{Theorem} \coqdocid{conjsg\_conj}: \ensuremath{\forall}\coqdocid{x} \coqdocid{y} \coqdocid{z}, \coqdocid{conjsg} (\coqdocid{x} \ensuremath{\times} \coqdocid{y}) \coqdocid{z} = \coqdocid{conjsg} \coqdocid{y} (\coqdocid{z}$^{\coqdocid{x}}$).\coqdoceol

\medskip
\noindent
\coqdockw{Theorem} \coqdocid{conjsg\_subgrp}: \ensuremath{\forall}\coqdocid{x}, \coqdocid{subgrp} (\coqdocid{conjsg} \coqdocid{x}).\coqdoceol

\medskip
\noindent
\coqdockw{Theorem} \coqdocid{conjsg\_image}: \ensuremath{\forall}\coqdocid{y},\coqdoceol
\coqdocindent{1.00em}
\coqdocid{conjsg} \coqdocid{y} =$_1$ \coqdocid{image} (\coqdocid{conjg} \coqdocid{y}$^{-1}$) \coqdocid{H}.\coqdoceol

\medskip
\noindent
\coqdockw{Theorem} \coqdocid{conjsg\_inv1}: \ensuremath{\forall}\coqdocid{x}, \coqdoceol
\coqdocindent{1.50em}
(\coqdocid{conjsg} \coqdocid{x}) =$_1$ \coqdocid{H} \ensuremath{\rightarrow} (\coqdocid{conjsg} \coqdocid{x}$^{-1}$) =$_1$ \coqdocid{H}.\coqdoceol

\medskip
\noindent
\coqdockw{Theorem} \coqdocid{conjsg\_card}: \ensuremath{\forall}\coqdocid{x},\coqdoceol
\coqdocindent{1.00em}
\coqdocid{card} (\coqdocid{conjsg} \coqdocid{x}) = \coqdocid{card} \coqdocid{H}.\coqdoceol

\medskip
\noindent
\coqdockw{Theorem} \coqdocid{conjsg\_subset}: \ensuremath{\forall}\coqdocid{x}, \coqdoceol
\coqdocindent{1.50em}
\coqdocid{subset} \coqdocid{H} (\coqdocid{conjsg} \coqdocid{x}) \ensuremath{\rightarrow} (\coqdocid{conjsg} \coqdocid{x}) =$_1$ \coqdocid{H}.\coqdoceol

\medskip
\noindent
\coqdockw{Theorem} \coqdocid{lcoset\_root}: \ensuremath{\forall}\coqdocid{x}, \coqdocid{lcoset} \coqdocid{H} \coqdocid{x} (\coqdocid{root} (\coqdocid{lcoset} \coqdocid{H}) \coqdocid{x}).\coqdoceol

\medskip
\noindent
\coqdockw{Definition} \coqdocid{normalb}:= \coqdocid{subset} \coqdocid{K} (\coqdocid{fun} \coqdocid{x} \ensuremath{\Rightarrow} \coqdocid{subset} \coqdocid{H} (\coqdocid{conjsg} \coqdocid{H} \coqdocid{x})).\coqdoceol

\medskip
\noindent
\coqdockw{Definition} \coqdocid{normal}: \coqdocid{Prop}:= \coqdocid{normalb}.\coqdoceol

\medskip
\noindent
\coqdockw{Hypothesis} \coqdocid{normal\_k}: \coqdocid{normal}.\coqdoceol

\medskip
\noindent
\coqdockw{Theorem} \coqdocid{conjsg\_normal}: \ensuremath{\forall}\coqdocid{x},
\coqdocid{K} \coqdocid{x} \ensuremath{\rightarrow} \coqdocid{conjsg} \coqdocid{x} =$_1$ \coqdocid{H}.\coqdoceol

\medskip
\noindent
\coqdockw{Definition} \coqdocid{rootSet}:= \coqdocid{subFin} (\coqdocid{setI} (\coqdocid{roots} (\coqdocid{lcoset} \coqdocid{H})) \coqdocid{K}).\coqdoceol

\medskip
\noindent
\coqdockw{Theorem} \coqdocid{card\_rootSet}: \coqdocid{card} \coqdocid{rootSet} = \coqdocid{lindex} \coqdocid{H} \coqdocid{K}.\coqdoceol

\medskip
\noindent
\coqdockw{Theorem} \coqdocid{unit\_root\_sub}:\coqdoceol
\coqdocindent{0.50em}
\coqdocid{setI} (\coqdocid{roots} (\coqdocid{lcoset} \coqdocid{H})) \coqdocid{K} (\coqdocid{root} (\coqdocid{lcoset} \coqdocid{H}) 1).\coqdoceol

\medskip
\noindent
\coqdockw{Definition} \coqdocid{unit\_root}: \coqdocid{rootSet}.\coqdoceol

\medskip
\noindent
\coqdockw{Definition} \coqdocid{mult\_root}: \coqdocid{rootSet} \ensuremath{\rightarrow} \coqdocid{rootSet} \ensuremath{\rightarrow} \coqdocid{rootSet}.\coqdoceol

\medskip
\noindent
\coqdockw{Definition} \coqdocid{inv\_root}: \coqdocid{rootSet} \ensuremath{\rightarrow} \coqdocid{rootSet}.\coqdoceol

\medskip
\noindent
\coqdockw{Theorem} \coqdocid{unitP\_root}: \ensuremath{\forall}\coqdocid{x}, \coqdocid{mult\_root} \coqdocid{unit\_root} \coqdocid{x} = \coqdocid{x}.\coqdoceol

\medskip
\noindent
\coqdockw{Theorem} \coqdocid{invP\_root}: \ensuremath{\forall}\coqdocid{x}, \coqdocid{mult\_root} (\coqdocid{inv\_root} \coqdocid{x}) \coqdocid{x} = \coqdocid{unit\_root}.\coqdoceol

\medskip
\noindent
\coqdockw{Theorem} \coqdocid{mulP\_root}: \ensuremath{\forall}\coqdocid{x$_1$} \coqdocid{x$_2$} \coqdocid{x$_3$}, \coqdoceol
\coqdocindent{1.00em}
\coqdocid{mult\_root} \coqdocid{x$_1$} (\coqdocid{mult\_root} \coqdocid{x$_2$} \coqdocid{x$_3$}) = \coqdocid{mult\_root} (\coqdocid{mult\_root} \coqdocid{x$_1$} \coqdocid{x$_2$}) \coqdocid{x$_3$}.\coqdoceol

\medskip
\noindent
\coqdockw{Definition} \coqdocid{root\_group}:= (\coqdocid{Group.Finite} \coqdocid{unitP\_root} \coqdocid{invP\_root} \coqdocid{mulP\_root}).\coqdoceol

\medskip
\noindent
\coqdockw{Theorem} \coqdocid{card\_root\_group}: \coqdocid{card} \coqdocid{root\_group} = \coqdocid{lindex} \coqdocid{H} \coqdocid{K}.\coqdoceol

\medskip
\noindent
\coqdockw{End} \coqdocid{Normal}.\coqdoceol

\medskip
\noindent
\coqdockw{Section} \coqdocid{NormalProp}.\coqdoceol

\medskip
\noindent
\coqdockw{Variables} (\coqdocid{G}: \coqdocid{finGroup}) (\coqdocid{H} \coqdocid{K}: \coqdocid{set} \coqdocid{G}).\coqdoceol
\noindent
\coqdockw{Hypothesis} \coqdocid{sgrp\_h}: \coqdocid{subgrp} \coqdocid{H}.\coqdoceol
\noindent
\coqdockw{Hypothesis} \coqdocid{sgrp\_k}: \coqdocid{subgrp} \coqdocid{K}.\coqdoceol
\noindent
\coqdockw{Hypothesis} \coqdocid{subset\_hk}: \coqdocid{subset} \coqdocid{H} \coqdocid{K}.\coqdoceol
\noindent
\coqdockw{Hypothesis} \coqdocid{normal\_hk}: \coqdocid{normal} \coqdocid{H} \coqdocid{K}.\coqdoceol

\medskip
\noindent
\coqdockw{Theorem} \coqdocid{normal\_subset}: \ensuremath{\forall}\coqdocid{L},\coqdoceol
\coqdocindent{1.00em}
\coqdocid{subgrp} \coqdocid{L} \ensuremath{\rightarrow} \coqdocid{subset} \coqdocid{H} \coqdocid{L} \ensuremath{\rightarrow} \coqdocid{subset} \coqdocid{L} \coqdocid{K} \ensuremath{\rightarrow} \coqdocid{normal} \coqdocid{H} \coqdocid{L}.\coqdoceol

\medskip
\noindent
\coqdockw{Definition} \coqdocid{RG}:= (\coqdocid{root\_group} \coqdocid{sgrp\_h} \coqdocid{sgrp\_k} \coqdocid{subset\_hk} \coqdocid{normal\_hk}).\coqdoceol

\medskip
\noindent
\coqdockw{Theorem} \coqdocid{th\_quotient}: \ensuremath{\forall}\coqdocid{x}, \coqdocid{K} \coqdocid{x} \ensuremath{\rightarrow}\coqdoceol
\coqdocindent{1.00em}
(\coqdocid{setI} (\coqdocid{roots} (\coqdocid{lcoset} \coqdocid{H})) \coqdocid{K} (\coqdocid{root} (\coqdocid{lcoset} \coqdocid{H}) \coqdocid{x})).\coqdoceol

\medskip
\noindent
\coqdockw{Definition} \coqdocid{quotient}: \coqdocid{G} \ensuremath{\rightarrow} \coqdocid{RG}.\coqdoceol

\medskip
\noindent
\coqdockw{Theorem} \coqdocid{quotient\_lcoset}: \ensuremath{\forall}\coqdocid{x}, \coqdocid{K} \coqdocid{x} \ensuremath{\rightarrow} \coqdocid{lcoset} \coqdocid{H} \coqdocid{x} (\coqdocid{val} (\coqdocid{quotient} \coqdocid{x})).\coqdoceol

\medskip
\noindent
\coqdockw{Theorem} \coqdocid{quotient1}: \ensuremath{\forall}\coqdocid{x}, \coqdocid{H} \coqdocid{x} \ensuremath{\rightarrow} \coqdocid{quotient} \coqdocid{x} = 1.\coqdoceol

\medskip
\noindent
\coqdockw{Theorem} \coqdocid{quotient\_morph}: \ensuremath{\forall}\coqdocid{x} \coqdocid{y},\coqdoceol
\coqdocindent{1.00em} \coqdocid{K} \coqdocid{x} \ensuremath{\rightarrow} \coqdocid{K} \coqdocid{y} \ensuremath{\rightarrow} \coqdocid{quotient}(\coqdocid{x} \ensuremath{\times} \coqdocid{y}) = \coqdocid{quotient}(\coqdocid{x}) \ensuremath{\times} \coqdocid{quotient}(\coqdocid{y}).\coqdoceol

\medskip
\noindent
\coqdockw{Theorem} \coqdocid{quotient\_image\_subgrp}: \ensuremath{\forall}\coqdocid{L},\coqdoceol
\coqdocindent{1.00em}
\coqdocid{subset} \coqdocid{H} \coqdocid{L} \ensuremath{\rightarrow} \coqdocid{subset} \coqdocid{L} \coqdocid{K} \ensuremath{\rightarrow} \coqdocid{subgrp} \coqdocid{L} \ensuremath{\rightarrow} \coqdocid{subgrp} (\coqdocid{image} \coqdocid{quotient} \coqdocid{L}).\coqdoceol

\medskip
\noindent
\coqdockw{Theorem} \coqdocid{quotient\_preimage\_subgrp}: \ensuremath{\forall}\coqdocid{L},\coqdoceol
\coqdocindent{1.00em}
\coqdocid{subgrp} \coqdocid{L} \ensuremath{\rightarrow} \coqdocid{subgrp} (\coqdocid{setI} (\coqdocid{preimage} \coqdocid{quotient} \coqdocid{L}) \coqdocid{K}).\coqdoceol

\medskip
\noindent
\coqdockw{Theorem} \coqdocid{quotient\_preimage\_subset\_h}: \ensuremath{\forall}\coqdocid{L},\coqdoceol
\coqdocindent{1.00em}
\coqdocid{subgrp} \coqdocid{L} \ensuremath{\rightarrow} \coqdocid{subset} \coqdocid{H} (\coqdocid{setI} (\coqdocid{preimage} \coqdocid{quotient} \coqdocid{L}) \coqdocid{K}).\coqdoceol

\medskip
\noindent
\coqdockw{Theorem} \coqdocid{quotient\_preimage\_subset\_k}: \ensuremath{\forall}\coqdocid{L},
\coqdocid{subset} (\coqdocid{setI} (\coqdocid{preimage} \coqdocid{quotient} \coqdocid{L}) \coqdocid{K}) \coqdocid{K}.\coqdoceol

\medskip
\noindent
\coqdockw{Theorem} \coqdocid{quotient\_index}: \ensuremath{\forall}\coqdocid{L},
\coqdocid{subset} \coqdocid{H} \coqdocid{L} \ensuremath{\rightarrow} \coqdocid{subset} \coqdocid{L} \coqdocid{K} \ensuremath{\rightarrow} \coqdocid{subgrp} \coqdocid{L} \ensuremath{\rightarrow} \coqdoceol
\coqdocindent{1.00em}\coqdocid{lindex} \coqdocid{H} \coqdocid{L} = \coqdocid{card} (\coqdocid{image} \coqdocid{quotient} \coqdocid{L}).\coqdoceol

\medskip
\noindent
\coqdockw{Theorem} \coqdocid{quotient\_image\_preimage}: \ensuremath{\forall}\coqdocid{L},\coqdoceol
\coqdocindent{1.00em}
\coqdocid{image} \coqdocid{quotient} (\coqdocid{setI} (\coqdocid{preimage} \coqdocid{quotient} \coqdocid{L}) \coqdocid{K}) =$_1$ \coqdocid{L}.\coqdoceol

\medskip
\noindent
\coqdockw{End} \coqdocid{NormalProp}.\coqdoceol

\medskip
\noindent
\coqdockw{Section} \coqdocid{Normalizer}.\coqdoceol

\medskip
\noindent
\coqdockw{Variables} (\coqdocid{G}: \coqdocid{finGroup}) (\coqdocid{H} \coqdocid{K}: \coqdocid{set} \coqdocid{G}).\coqdoceol

\medskip
\noindent
\coqdockw{Hypothesis} \coqdocid{sgrp\_h}: \coqdocid{subgrp} \coqdocid{H}.\coqdoceol
\noindent
\coqdockw{Hypothesis} \coqdocid{sgrp\_k}: \coqdocid{subgrp} \coqdocid{K}.\coqdoceol
\noindent
\coqdockw{Hypothesis} \coqdocid{subset\_hk}: \coqdocid{subset} \coqdocid{H} \coqdocid{K}.\coqdoceol

\medskip
\noindent
\coqdockw{Definition} \coqdocid{normaliser} \coqdocid{x}:= \coqdoceol
\coqdocindent{1.00em}
(\coqdocid{subset}  \coqdocid{K} (\coqdocid{fun} \coqdocid{z} \ensuremath{\Rightarrow} (\coqdocid{conjsg} \coqdocid{x} \coqdocid{z} =${_d}$ \coqdocid{H} \coqdocid{z}))) \&\& \coqdocid{K} \coqdocid{x}.\coqdoceol

\medskip
\noindent
\coqdockw{Theorem} \coqdocid{normaliser\_grp}: \coqdocid{subgrp} \coqdocid{normaliser}.\coqdoceol

\medskip
\noindent
\coqdockw{Theorem} \coqdocid{normaliser\_subset}: \coqdocid{subset} \coqdocid{normaliser} \coqdocid{K}.\coqdoceol

\medskip
\noindent
\coqdockw{Theorem} \coqdocid{subset\_normaliser}: \coqdocid{subset} \coqdocid{H} \coqdocid{normaliser}.\coqdoceol

\medskip
\noindent
\coqdockw{Theorem} \coqdocid{normaliser\_normal}: \coqdocid{normal} \coqdocid{H} \coqdocid{normaliser}.\coqdoceol

\medskip
\noindent
\coqdockw{Theorem} \coqdocid{card\_normaliser}:\coqdoceol
\noindent
\coqdocid{card} (\coqdocid{root\_group} \coqdocid{sgrp\_h} \coqdocid{normaliser\_grp} \coqdocid{subset\_normaliser}\coqdoceol
\coqdocindent{6.00em}
\coqdocid{normaliser\_normal}) = \coqdocid{lindex} \coqdocid{H} \coqdocid{normaliser}.\coqdoceol

\medskip
\noindent
\coqdockw{End} \coqdocid{Normalizer}.\coqdoceol

\medskip
\noindent
\coqdockw{Section} \coqdocid{Eq}.\coqdoceol

\medskip
\noindent
\coqdockw{Variables} \coqdocid{G}: \coqdocid{finGroup}.\coqdoceol

\medskip
\noindent
\coqdockw{Theorem} \coqdocid{eq\_conjsg}: \ensuremath{\forall}\coqdocid{a} \coqdocid{b} \coqdocid{x}, \coqdocid{a} =$_1$ \coqdocid{b} \ensuremath{\rightarrow} \coqdocid{conjsg} \coqdocid{a} \coqdocid{x} =$_1$ \coqdocid{conjsg} \coqdocid{b} \coqdocid{x}.\coqdoceol

\medskip
\noindent
\coqdockw{End} \coqdocid{Eq}.\coqdoceol

\medskip
\noindent
\coqdockw{Section} \coqdocid{Root}.\coqdoceol

\medskip
\noindent
\coqdockw{Variable} (\coqdocid{G}: \coqdocid{finGroup}) (\coqdocid{H}: \coqdocid{set} \coqdocid{G}).\coqdoceol

\medskip
\noindent
\coqdockw{Hypothesis} \coqdocid{sgrp\_h}: \coqdocid{subgrp} \coqdocid{H}.\coqdoceol

\medskip
\noindent
\coqdockw{Theorem} \coqdocid{root\_lcoset1}: \coqdocid{H} (\coqdocid{root} (\coqdocid{lcoset} \coqdocid{H}) 1).\coqdoceol

\medskip
\noindent
\coqdockw{Theorem} \coqdocid{root\_lcosetd}: \ensuremath{\forall}\coqdocid{a}, \coqdocid{H} (\coqdocid{a}$^{-1}$ \ensuremath{\times} \coqdocid{root} (\coqdocid{lcoset} \coqdocid{H}) \coqdocid{a}).\coqdoceol

\medskip
\noindent
\coqdockw{End} \coqdocid{Root}.\coqdoceol

\coqdocmodule{leftTranslation}

\medskip
\noindent
\coqdockw{Section} \coqdocid{LeftTrans}.\coqdoceol

\medskip
\noindent
\coqdockw{Variable} (\coqdocid{G}: \coqdocid{finGroup}) (\coqdocid{H} \coqdocid{K} \coqdocid{L}: \coqdocid{set} \coqdocid{G}).\coqdoceol

\medskip
\noindent
\coqdockw{Hypothesis} \coqdocid{sgrp\_k}: \coqdocid{subgrp} \coqdocid{K}.\coqdoceol
\noindent
\coqdockw{Hypothesis} \coqdocid{sgrp\_l}: \coqdocid{subgrp} \coqdocid{L}.\coqdoceol
\noindent
\coqdockw{Hypothesis} \coqdocid{sgrp\_h}: \coqdocid{subgrp} \coqdocid{H}.\coqdoceol
\noindent
\coqdockw{Hypothesis} \coqdocid{subset\_hk}: \coqdocid{subset} \coqdocid{H} \coqdocid{K}.\coqdoceol
\noindent
\coqdockw{Hypothesis} \coqdocid{subset\_lk}: \coqdocid{subset} \coqdocid{L} \coqdocid{K}.\coqdoceol

\medskip
\noindent
\coqdockw{Definition} \coqdocid{ltrans}: \coqdocid{G} \ensuremath{\rightarrow} \coqdocid{rootSet} \coqdocid{L} \coqdocid{K} \ensuremath{\rightarrow} \coqdocid{rootSet} \coqdocid{L} \coqdocid{K}.\coqdoceol

\medskip
\noindent
\coqdockw{Theorem} \coqdocid{ltrans\_bij}: \ensuremath{\forall}\coqdocid{x},
\coqdocid{H} \coqdocid{x} \ensuremath{\rightarrow} \coqdocid{bijective} (\coqdocid{ltrans} \coqdocid{x}).\coqdoceol

\medskip
\noindent
\coqdockw{Theorem} \coqdocid{ltrans\_morph}: \ensuremath{\forall}\coqdocid{x} \coqdocid{y} \coqdocid{z},\coqdoceol
\coqdocindent{1.00em}
\coqdocid{H} \coqdocid{x} \ensuremath{\rightarrow} \coqdocid{H} \coqdocid{y} \ensuremath{\rightarrow} \coqdocid{ltrans} (\coqdocid{x} \ensuremath{\times} \coqdocid{y}) \coqdocid{z} = \coqdocid{ltrans} \coqdocid{x} (\coqdocid{ltrans} \coqdocid{y} \coqdocid{z}).\coqdoceol

\medskip
\noindent
\coqdockw{End} \coqdocid{LeftTrans}.\coqdoceol

\coqdocmodule{sylow}

\medskip
\noindent
\coqdockw{Section} \coqdocid{Cauchy}.\coqdoceol

\medskip
\noindent
\coqdockw{Variable} (\coqdocid{G}: \coqdocid{finGroup}) (\coqdocid{H}: \coqdocid{set} \coqdocid{G}).\coqdoceol
\noindent
\coqdockw{Hypothesis} \coqdocid{sgrp\_h}: \coqdocid{subgrp} \coqdocid{H}.\coqdoceol

\medskip
\noindent
\coqdockw{Variable} \coqdocid{p}: \coqdocid{nat}.\coqdoceol
\noindent
\coqdockw{Hypothesis} \coqdocid{prime\_p}: \coqdocid{prime} \coqdocid{p}.\coqdoceol
\noindent
\coqdockw{Hypothesis} \coqdocid{p\_divides\_h}: \coqdocid{p} \ensuremath{|} \coqdocid{card} \coqdocid{H}.\coqdoceol

\medskip
\noindent
\coqdockw{Theorem} \coqdocid{cauchy}: \ensuremath{\exists} \coqdocid{a},\coqdocid{H} \coqdocid{a} \&\& \coqdocid{card} (\coqdocid{cyclic} \coqdocid{a}) =${_d}$ \coqdocid{p}.\coqdoceol

\medskip
\noindent
\coqdockw{End} \coqdocid{Cauchy}.\coqdoceol

\medskip
\noindent
\coqdockw{Section} \coqdocid{Sylow}.\coqdoceol

\medskip
\noindent
\coqdockw{Variable} (\coqdocid{G}: \coqdocid{finGroup}) (\coqdocid{K}: \coqdocid{set} \coqdocid{G}).\coqdoceol
\noindent
\coqdockw{Hypothesis} \coqdocid{sgrp\_k}: \coqdocid{subgrp} \coqdocid{K}.\coqdoceol

\medskip
\noindent
\coqdockw{Variable} \coqdocid{p}: \coqdocid{nat}.\coqdoceol
\noindent
\coqdockw{Hypothesis} \coqdocid{prime\_p}: \coqdocid{prime} \coqdocid{p}.\coqdoceol

\medskip
\noindent
\coqdockw{Let} \coqdocid{n}:= \coqdocid{dlogn} \coqdocid{p} (\coqdocid{card} \coqdocid{K}).\coqdoceol

\medskip
\noindent
\coqdockw{Hypothesis} \coqdocid{n\_pos}: 0 < \coqdocid{n}.\coqdoceol

\medskip
\noindent
\coqdockw{Definition} \coqdocid{sylow} \coqdocid{L}:= (\coqdocid{subgrpb} \coqdocid{L}) \&\& (\coqdocid{subset} \coqdocid{L} \coqdocid{K}) \&\& (\coqdocid{card} \coqdocid{L} =${_d}$ \coqdocid{p}$^{\coqdocid{n}}$).\coqdoceol

\medskip
\noindent
\coqdockw{Theorem} \coqdocid{eq\_sylow}: \ensuremath{\forall}\coqdocid{a} \coqdocid{b}, \coqdocid{a} =$_1$ \coqdocid{b} \ensuremath{\rightarrow} \coqdocid{sylow} \coqdocid{a} = \coqdocid{sylow} \coqdocid{b}.\coqdoceol

\medskip
\noindent
\coqdockw{Theorem} \coqdocid{sylow\_conjsg}: \ensuremath{\forall}\coqdocid{L$_1$} \coqdocid{x}, \coqdocid{K} \coqdocid{x} \ensuremath{\rightarrow}
\coqdocid{sylow} \coqdocid{L$_1$} \ensuremath{\rightarrow} \coqdocid{sylow} (\coqdocid{conjsg} \coqdocid{L$_1$} \coqdocid{x}).\coqdoceol

\medskip
\noindent
\coqdockw{Theorem} \coqdocid{sylow1\_rec}: \ensuremath{\forall}\coqdocid{i} \coqdocid{Hi}, 0 < \coqdocid{i} \ensuremath{\rightarrow} \coqdocid{i} < \coqdocid{n} \ensuremath{\rightarrow} \coqdoceol
\coqdocindent{1.00em}
\coqdocid{subgrp} \coqdocid{Hi} \ensuremath{\rightarrow} \coqdocid{subset} \coqdocid{Hi} \coqdocid{K} \ensuremath{\rightarrow} \coqdocid{card} \coqdocid{Hi} = \coqdocid{p}$^{\coqdocid{i}}$ \ensuremath{\rightarrow}\coqdoceol
\coqdocindent{1.00em}
\ensuremath{\exists} \coqdocid{H}: \coqdocid{set} \coqdocid{G},\coqdoceol
\coqdocindent{1.50em} \coqdocid{subgrp} \coqdocid{H} \ensuremath{\land} \coqdocid{subset} \coqdocid{Hi} \coqdocid{H} \ensuremath{\land} \coqdocid{subset} \coqdocid{H} \coqdocid{K} 
\ensuremath{\land} \coqdocid{normal} \coqdocid{Hi} \coqdocid{H} \ensuremath{\land} \coqdocid{card} \coqdocid{H} = \coqdocid{p}$^{\coqdocid{i} \ensuremath{+} 1}$.\coqdoceol

\medskip
\noindent
\coqdockw{Theorem} \coqdocid{sylow1}: \ensuremath{\forall}\coqdocid{i}, 0 < \coqdocid{i} \ensuremath{\rightarrow} \coqdocid{i} \ensuremath{\le} \coqdocid{n} \ensuremath{\rightarrow} \coqdoceol
\coqdocindent{1.00em}
\ensuremath{\exists} \coqdocid{H}: \coqdocid{set} \coqdocid{G}, \coqdocid{subgrp} \coqdocid{H} \ensuremath{\land} \coqdocid{subset} \coqdocid{H} \coqdocid{K} \ensuremath{\land} \coqdocid{card} \coqdocid{H} = \coqdocid{p}$^{\coqdocid{i}}$.\coqdoceol

\medskip
\noindent
\coqdockw{Theorem} \coqdocid{sylow1\_cor}: \ensuremath{\exists} \coqdocid{H}: \coqdocid{set} \coqdocid{G}, \coqdocid{sylow} \coqdocid{H}.\coqdoceol

\medskip
\noindent
\coqdockw{Theorem} \coqdocid{sylow2}: \ensuremath{\forall}\coqdocid{H} \coqdocid{L} \coqdocid{i},0 <\coqdocid{i} \ensuremath{\rightarrow} \coqdocid{i} \ensuremath{\le} \coqdocid{n} \ensuremath{\rightarrow}\coqdoceol
\coqdocindent{0.50em}
\coqdocid{subgrp} \coqdocid{H} \ensuremath{\rightarrow} \coqdocid{subset} \coqdocid{H} \coqdocid{K} \ensuremath{\rightarrow}  \coqdocid{card} \coqdocid{H} = \coqdocid{p$^i$}  \ensuremath{\rightarrow} \coqdocid{sylow} \coqdocid{L} \ensuremath{\rightarrow}\coqdoceol
\coqdocindent{1.00em}
\ensuremath{\exists} \coqdocid{x}, (\coqdocid{K} \coqdocid{x}) \&\& \coqdocid{subset} \coqdocid{H} (\coqdocid{conjsg} \coqdocid{L} \coqdocid{x}).\coqdoceol

\medskip
\noindent
\coqdockw{Theorem} \coqdocid{sylow2\_cor}: \ensuremath{\forall}\coqdocid{L$_1$} \coqdocid{L$_2$},
\coqdocid{sylow} \coqdocid{L$_1$} \ensuremath{\rightarrow} \coqdocid{sylow} \coqdocid{L$_2$} \ensuremath{\rightarrow}\coqdoceol
\coqdocindent{1.00em} \ensuremath{\exists} \coqdocid{x}, (\coqdocid{K} \coqdocid{x}) \ensuremath{\land} (\coqdocid{L$_2$} =$_1$ \coqdocid{conjsg} \coqdocid{L$_1$} \coqdocid{x}).\coqdoceol

\medskip
\noindent
\coqdockw{Definition} \coqdocid{syset} \coqdocid{p}:= \coqdocid{sylow} (\coqdocid{val} \coqdocid{p}).\coqdoceol

\medskip
\noindent
\coqdockw{Theorem} \coqdocid{sylow3\_div}: \coqdocid{card} \coqdocid{syset} \ensuremath{|} \coqdocid{card} \coqdocid{K}.\coqdoceol

\medskip
\noindent
\coqdockw{End} \coqdocid{Sylow}.\coqdoceol

\medskip
\noindent
\coqdockw{Section} \coqdocid{SylowAux}.\coqdoceol

\medskip
\noindent
\coqdockw{Variable} (\coqdocid{G}: \coqdocid{finGroup}) (\coqdocid{H} \coqdocid{K} \coqdocid{L}: \coqdocid{set} \coqdocid{G}).\coqdoceol
\noindent
\coqdockw{Hypothesis} \coqdocid{sgrp\_k}: \coqdocid{subgrp} \coqdocid{K}.\coqdoceol
\noindent
\coqdockw{Hypothesis} \coqdocid{sgrp\_l}: \coqdocid{subgrp} \coqdocid{L}.\coqdoceol
\noindent
\coqdockw{Hypothesis} \coqdocid{sgrp\_h}: \coqdocid{subgrp} \coqdocid{H}.\coqdoceol
\noindent
\coqdockw{Hypothesis} \coqdocid{subset\_hl}: \coqdocid{subset} \coqdocid{H} \coqdocid{L}.\coqdoceol
\noindent
\coqdockw{Hypothesis} \coqdocid{subset\_lk}: \coqdocid{subset} \coqdocid{L} \coqdocid{K}.\coqdoceol

\medskip
\noindent
\coqdockw{Variable} \coqdocid{p}: \coqdocid{nat}.\coqdoceol
\noindent
\coqdockw{Hypothesis} \coqdocid{prime\_p}: \coqdocid{prime} \coqdocid{p}.\coqdoceol
\noindent
\coqdockw{Let} \coqdocid{n}:= \coqdocid{dlogn} \coqdocid{p} (\coqdocid{card} \coqdocid{K}).\coqdoceol
\noindent
\coqdockw{Hypothesis} \coqdocid{n\_pos}: 0 < \coqdocid{n}.\coqdoceol

\medskip
\noindent
\coqdockw{Theorem} \coqdocid{sylow\_subset}: \coqdocid{sylow} \coqdocid{K} \coqdocid{p} \coqdocid{H} \ensuremath{\rightarrow} \coqdocid{sylow} \coqdocid{L} \coqdocid{p} \coqdocid{H}.\coqdoceol

\medskip
\noindent
\coqdockw{End} \coqdocid{SylowAux}.\coqdoceol

\medskip
\noindent
\coqdockw{Section} \coqdocid{Sylow3}.\coqdoceol

\medskip
\noindent
\coqdockw{Variable} (\coqdocid{G}: \coqdocid{finGroup}) (\coqdocid{K}: \coqdocid{set} \coqdocid{G}).\coqdoceol
\noindent
\coqdockw{Hypothesis} \coqdocid{sgrp\_k}: \coqdocid{subgrp} \coqdocid{K}.\coqdoceol

\medskip
\noindent
\coqdockw{Variable} \coqdocid{p}: \coqdocid{nat}.\coqdoceol
\noindent
\coqdockw{Hypothesis} \coqdocid{prime\_p}: \coqdocid{prime} \coqdocid{p}.\coqdoceol

\medskip
\noindent
\coqdockw{Let} \coqdocid{n}:= \coqdocid{dlogn} \coqdocid{p} (\coqdocid{card} \coqdocid{K}).\coqdoceol

\medskip
\noindent
\coqdockw{Hypothesis} \coqdocid{n\_pos}: 0 < \coqdocid{n}.\coqdoceol

\medskip
\noindent
\coqdockw{Theorem} \coqdocid{sylow3\_mod}: \coqdocid{card} (\coqdocid{syset} \coqdocid{K} \coqdocid{p}) \% \coqdocid{p} = 1.\coqdoceol

\medskip
\noindent
\coqdockw{End} \coqdocid{Sylow3}.\coqdoceol

\end{document}